\documentclass[11pt]{article}

\usepackage[T1]{fontenc}
\usepackage[utf8]{inputenc}
\usepackage{lmodern}
\usepackage{microtype}
\usepackage[a4paper,margin=1.02in]{geometry}
\usepackage{setspace}
\usepackage{amsmath,amssymb,amsthm,mathtools,bm}
\usepackage{booktabs,array,tabularx,multirow}
\usepackage{enumitem}
\usepackage{graphicx}
\graphicspath{{./}{figures/}}
\usepackage{natbib}
\usepackage[colorlinks=true,allcolors=blue!55!black]{hyperref}
\hypersetup{
 pdftitle={Robustness or Crowding: Experimental Design for Trading Strategy Capacity},
 pdfauthor={Alejandro Rodriguez Dominguez and Miquel Noguer i Alonso},
 pdfkeywords={experimental design; switchback; carryover; treatment persistence; strategy capacity; arbitrage capital}
}
\usepackage[nameinlink,capitalise,noabbrev]{cleveref}
\crefname{assumption}{Assumption}{Assumptions}
\Crefname{assumption}{Assumption}{Assumptions}
\crefname{definition}{Definition}{Definitions}
\Crefname{definition}{Definition}{Definitions}
\crefname{remark}{Remark}{Remarks}
\Crefname{remark}{Remark}{Remarks}
\crefname{proposition}{Proposition}{Propositions}
\Crefname{proposition}{Proposition}{Propositions}
\crefname{corollary}{Corollary}{Corollaries}
\Crefname{corollary}{Corollary}{Corollaries}
\crefname{lemma}{Lemma}{Lemmas}
\Crefname{lemma}{Lemma}{Lemmas}
\usepackage{xcolor}
\usepackage{tikz}
\usetikzlibrary{arrows.meta,positioning,calc,decorations.pathreplacing}
\usepackage{float}
\usepackage{placeins}
\usepackage{caption}
\usepackage{subcaption}
\usepackage{url}

\newtheorem{theorem}{Theorem}[section]
\newtheorem{proposition}[theorem]{Proposition}

\newtheorem{corollary}[theorem]{Corollary}
\newtheorem{assumption}[theorem]{Assumption}
\newtheorem{definition}[theorem]{Definition}
\newtheorem{remark}[theorem]{Remark}

\newcommand{\E}{\mathbb{E}}
\newcommand{\Prob}{\mathbb{P}}
\newcommand{\Var}{\operatorname{Var}}
\newcommand{\Cov}{\operatorname{Cov}}

\newcommand{\norm}[1]{\left\lVert #1 \right\rVert}

\newcommand{\Hcal}{\mathcal{H}}
\newcommand{\Fcal}{\mathcal{F}}

\newcommand{\bown}[1][]{\beta^{\mathrm{cap}}_{\mathrm{own}#1}}
\newcommand{\btot}[1][]{\beta^{\mathrm{cap}}_{\mathrm{strat}#1}}

\title{\textbf{Robustness or Crowding:}\\
\textbf{Experimental Design for Trading Strategy Capacity}}
\author{Alejandro Rodr{\'i}guez Dom{\'i}nguez\thanks{Department of Quantitative Analysis and Artificial Intelligence, Miralta Finance Bank, Madrid, and Department of Computer Science, University of Reading. Corresponding author. The views expressed are those of the authors and do not represent the views of any affiliated institution.} \and
Miquel Noguer i Alonso\thanks{Artificial Intelligence Finance Institute and Courant Institute of Mathematical Sciences, New York University.}}

\date{5 August 2026}

\begin{document}
\maketitle

\begin{abstract}
\noindent How much capital a trading strategy can absorb before its edge disappears is a causal question about how much is deployed, but it is answered with observational proxies that rest on incompatible assumptions. We ask what experiment would answer it instead, and show that two features of the problem interact to constrain any answer. Deployed capital erodes the edge gradually, so a trial of fixed length measures less than the eventual effect; and parallel implementations of one strategy trade the same securities, so they are not independent units. Comparing implementations on the same date removes market-wide shocks, which is what makes the comparison credible. But the crowding created by the strategy's own accumulated position is common to those implementations too, and an arbitrary date effect absorbs it exactly: the comparison that makes the experiment robust is the one that prevents it from measuring the crowding capacity is about. A same-date design recovers one implementation's private response at the prevailing level of aggregate positioning, and reaching the aggregate effect requires either implementations with deliberately different exposure to that position or variation in it over time. We characterise what each route identifies and what it costs, establish how far a fixed holding period understates the eventual effect and how to correct for it, and show what a finite set of deployment levels can and cannot reveal. A calibration on a purpose-built panel illustrates the resulting design rules and prices a study that would follow them.
\end{abstract}


\section{Introduction}

How much capital a trading strategy can absorb before its edge is exhausted is a question about what would happen at levels of deployment other than the one realised. It is therefore causal, and the observational route to it has not converged. Deployed arbitrage capital has been inferred from the cross-section of short interest \citep{hanson2014}, from return comovement among the constituents of a strategy \citep{lou2022}, from institutional holdings and their liquidity footprint \citep{brown2022crowded,chincarini2026}, and from persistent flows into the funds that run these strategies \citep{dong2025}. These measures rest on different primitives, they are validated by their correlation with subsequent returns, and they have rarely been benchmarked against one another. That returns fall as capital accumulates is not in doubt: it organises the equilibrium model of \citet{berkgreen2004}, it is estimated for funds by \citet{pastor2015scale}, and it is the standard reading of the post-publication decay documented by \citet{mcleanpontiff2016} and replicated by \citet{houxuezhang2020} and \citet{chenzimmermann2022}. What is missing is the counterfactual that a capacity statement requires.

\begin{figure}[t]
\centering
\begin{tikzpicture}[>=Latex,font=\small,
  arm/.style={draw,rounded corners=2pt,minimum width=32mm,minimum height=6mm,align=center,inner sep=2pt},
  lab/.style={font=\footnotesize,align=center}]
  \node[lab] (sl) at (-3.3,0) {comparable\\sleeves};
  \node[arm] (a0) at (0,1.15) {control, scale $\beta_0$};
  \node[arm] (a1) at (0,0.39) {arm 1, scale $\beta_1$};
  \node[arm] (a2) at (0,-0.39) {arm 2, scale $\beta_2$};
  \node[arm] (a3) at (0,-1.15) {arm 3, scale $\beta_3$};
  \draw[dashed,gray!65] (-1.85,1.58) rectangle (1.85,-1.58);
  \node[lab,anchor=south,gray!50!black] at (0,1.64) {randomised within the same calendar dates, held $L$ periods};
  \foreach \n in {a0,a1,a2,a3} \draw[->,gray!65] (sl) -- (\n.west);
  \node[lab] (out) at (4.2,0) {adjusted return\\of each arm};
  \foreach \n in {a0,a1,a2,a3} \draw[->,gray!65] (\n.east) -- (out);
  \node[lab] (con) at (4.2,-2.75) {within-date contrast\\$\widehat D_t$};
  \node[lab] (cur) at (0.1,-2.75) {finite-hold curve\\$m^{\mathrm{avg}}_{L}(\beta)$};
  \node[lab] (cap) at (-3.3,-2.75) {steady state\\and own-sleeve capacity};
  \draw[->,gray!65] (out) -- (con);
  \draw[->,gray!65] (con) -- (cur);
  \draw[->,gray!65] (cur) -- (cap);
\end{tikzpicture}
\caption{The capacity experiment. Parallel sleeves running one strategy are randomly assigned deployment scales $\beta_0<\beta_1<\beta_2<\beta_3$, where $\beta_0=0$ is the control arm and the rest are multiples of a reference book; each assignment is held for $L$ consecutive periods, called a block. The outcome is the factor-adjusted return, and $\widehat D_t$ denotes the contrast between a treated and a control sleeve on the same date, so that anything shared by the sleeves in that period differences out. Averaging the contrast over the block gives $m^{\mathrm{avg}}_L(\beta)$, the finite-hold capacity curve, which understates the steady-state curve because the erosion stock has not finished accumulating when the block ends; $G_L\in(0,1]$ is the fraction that has accumulated on average over the block, so dividing by it carries the arm contrast from the finite-hold effect to the steady-state effect under the maintained dynamics. The intercept and the component of erosion common to all sleeves are not scaled by $G_L$, and the estimand throughout is the own-sleeve capacity curve traced at a fixed average deployment.}
\label{fig:experiment}
\end{figure}
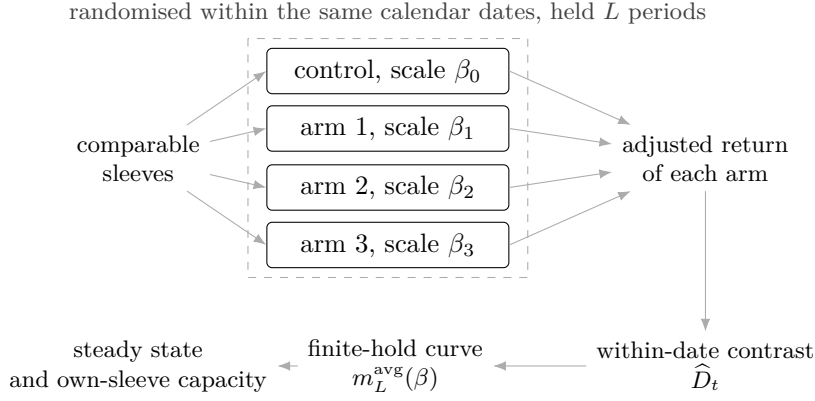

This paper takes the interventional route and designs one experiment, set out in \Cref{fig:experiment}. Parallel implementations of a strategy, which we call sleeves, are randomly assigned to different deployment scales; each assignment is held for a fixed number of periods; and the arms are compared within the same calendar dates. Because sleeves trade the same securities they are not isolated units, so we restrict how they interfere rather than assuming they do not, and the estimand is the effect of one sleeve's own assignment holding average deployment fixed, which is not the effect of scaling the whole strategy. Four questions then arise, and they organise everything that follows. What does such an experiment identify, and how does that depend on how long each assignment is held? How should the outcome recorded over a block be turned into an estimate, and is there a best way of doing so? What determines how the requirement scales when the experiment is replicated across sleeves? And how much of the answer rests on quantities the experiment itself cannot measure?

The answers turn on one feature of the setting. Capital deployed against a signal accumulates into a stock that dissipates slowly, a persistence visible in the price impact literature, where impact is concave in size \citep{toth2011,sato2024} and decays slowly and only partially \citep{brokmann2015,bucci2019}. Because erosion accumulates, what a design measures depends on how long the assigned level has been held: the shortest window, which period-by-period randomisation implements, identifies only the impact-period effect, under a tenth of the steady-state slope at the persistence calibrated below.

The contributions are of four kinds, and they lie in the design rather than in the apparatus: the tools are standard and are named where they are used. On identification, the accumulation kernel is characterised rather than assumed, capacity on a finite grid of arms is set identified with a sharp set open at its upper endpoint, and the within-block trajectory identifies the shape of the kernel without parametric restriction while bounding the steady-state effect from one side. On attenuation and recovery, each way of sampling the block induces its own attenuation factor, so the choice of outcome is a choice of estimand, and among unbiased linear recoveries the minimum-variance weights are explicit. On replication, robustness to arbitrary calendar shocks is a rank condition and cancellation of a common component is a zero-sum condition on the contrast weights; under homogeneous overlap that same condition removes the aggregate crowding, so calendar robustness and aggregate identification are mutually exclusive, and efficiency is a constrained least-squares problem. On cost, the experiment is priced in forgone edge rather than calendar time, the local opportunity cost is bounded away from zero, and an impact model estimated on execution records bounds capacity from above and never from both sides, the wedge being what the design measures. Two of the four inputs are transported under bridges that are stated and not testable here, and the calendar requirement is more sensitive to those bridges than to anything estimated on the panel.

One result subsumes the others as a design statement. An unrestricted calendar component is observationally equivalent to one that has absorbed the aggregate crowding, so that component is not identified from within-date variation by any estimator; the familiar statement that a zero-sum contrast removes both is the corollary for linear contrasts. Robustness to calendar shocks and identification of aggregate crowding are therefore mutually exclusive in any within-date design, and a contemporaneous experiment identifies a conditional marginal effect at low cost rather than the capacity of the strategy. Reaching the latter means either accepting the calendar exposure the design was built to avoid, at a cost we compute, or engineering heterogeneous overlap between sleeves, which recovers the second estimand at the price of a weaker exposure mapping.

Two further results matter in the opposite direction from the one they invite. The minimum-variance weights require the accumulation kernel, and in our equal-resource experiments, estimating that kernel inside the experiment costs more in variance than transporting a wrong one costs in bias, except when the transported form is badly misspecified, so the gain they promise is attainable only when the bridge is far from the truth. A sharp identified set is also not a confidence set: the interval formed between two estimated arm means covers the true capacity less than half the time, and sequential refinement makes that worse rather than better, so where arms are placed and what is reported have to be decided separately.

\begin{figure}[t]
\centering
\includegraphics[width=.92\textwidth]{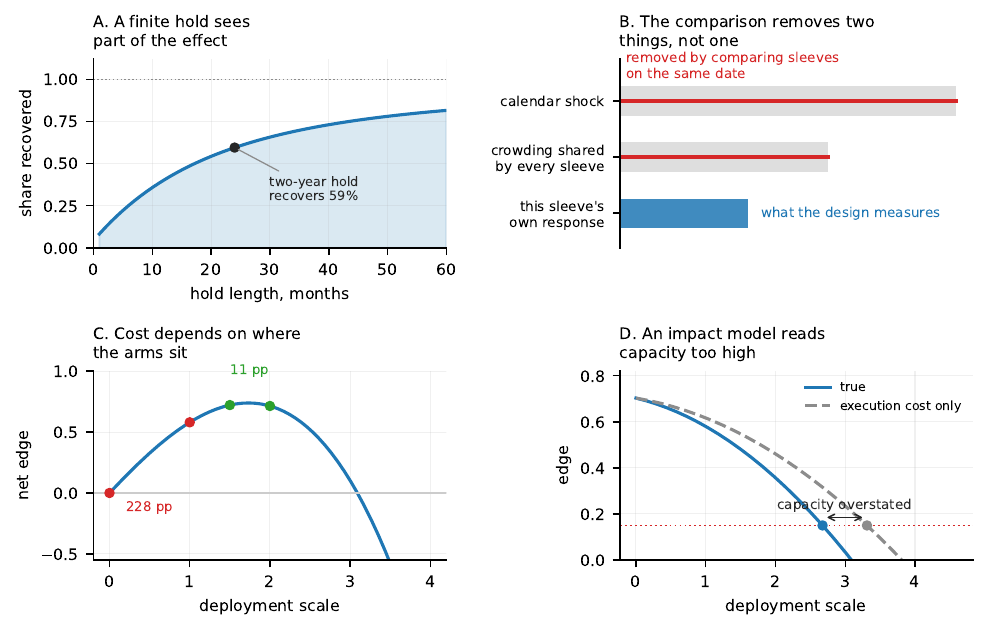}
\caption{The argument in four panels. \textbf{A.} Erosion accumulates, so a block of finite length sees only part of the steady-state effect: at the calibrated speed of accumulation a two-year hold recovers about three fifths of it, and the shortfall has to be corrected before the number means capacity. \textbf{B.} Comparing sleeves on the same date removes whatever they share in that period. The calendar shock is one such thing, which is why the comparison is robust; the crowding generated by the strategy's aggregate position is another, which is why the design measures a sleeve's own response and not the strategy's. \textbf{C.} Net edge is flat near the scale a book would run, so an arm placed there costs almost nothing while still separating the two arms. Comparing a zero arm with a unit arm gives up $228$ cumulative percentage points of edge; arms at $1.5$ and $2.0$ take essentially as long and give up $11$. \textbf{D.} An impact model fitted to execution records prices the cost of trading and not the decay of the edge, so it locates the crossing with the exit hurdle too far to the right and reports capacity that is too high.}
\label{fig:overview}
\end{figure}

\Cref{fig:overview} summarises the four findings; the sections that follow establish them. The paper proceeds as follows. Section 2 places the design in the literatures it draws on. Section 3 develops the theory: it sets out the experiment, its assumptions and its estimands; derives the attenuation each sampling rule induces and the minimum-variance recovery of the steady-state effect; establishes what determines the scale of a study and what a within-date contrast can and cannot identify; prices the experiment in forgone edge and bounds what an execution-cost model can deliver; and treats the separate design that detects path dependence. Section 4 takes those results to data: it calibrates the inputs on a panel built for the purpose, validates the attenuation and replication results by simulation, prices the resulting study, compares it against the impact models desks already run, and sets out which design to choose. No deployment experiment is run. Section 5 concludes, and the appendices hold the supporting results, the derivations, the ramp design, the outcome convention, the pre-registration protocol and the simulation details. Proofs not given in place, together with the details of the calibration and of the simulations, are in the appendices.

\section{Related literature}

The design we study is a switchback: a single unit is exposed to a sequence of randomly assigned policies, and the carryover between blocks governs what it identifies. \citet{bojinov2023} give the design theory for that setting and \citet{bojinovshephard2019} the randomisation-test treatment of time-series experiments in trading; the estimand here differs in that the carryover is the object of interest rather than a nuisance. For simultaneous inference across arms we draw on \citet{cck2014}, for the g-computation identity on \citet{robins1986}, and for the exposure mapping on \citet{hudgenshalloran2008} and \citet{aronowsamii2017}.

The scale of deployment is a continuous and sequentially assigned exposure, so the wider econometrics is that of continuous and dynamic treatments. The generalized propensity score of \citet{hiranoimbens2004} and the developments of \citet{kennedy2017}, \citet{hudson2024} and \citet{zhangchen2025} supply the estimand and its inference, while \citet{schomaker2024} treat continuous interventions at multiple time points and the positivity failures they generate. We work with finitely many arms, which avoids those failures and makes the carryover explicit, at the cost of identifying the curve only on the assigned grid. Interference is economic rather than social, and the exposure mapping used below is the device of \citet{hudgenshalloran2008} and \citet{aronowsamii2017}, which reduces a potential outcome indexed by the whole assignment vector to one indexed by a low-dimensional summary; the warning against conditioning on quantities affected by treatment is due to \citet{rosenbaum1984}, and covariate adjustment under randomisation follows \citet{lin2013}.

The persistence that drives our results is itself documented. Price impact is concave in size, following the square-root law of \citet{toth2011} confirmed across venues by \citet{sato2024} and surveyed in \citet{bouchaud2018}, and it decays slowly and only partially \citep{brokmann2015,bucci2019}. That partial decay is the empirical content of the stock recursion we assume.

The adjustment set is borrowed rather than devised. \citet{rodriguez2023} introduces the commonality principle as a criterion for selecting a portfolio's common drivers and embeds its constituents in the resulting sensitivity space, and \citet{rodriguez2025causal} derives that principle from Reichenbach's common cause condition and reads the sensitivities as causal objects. We use the driver screen and the metric; no design result below depends on the causal reading.

Generalised least squares under a linear restriction, the Frisch--Waugh--Lovell rank condition, impulse-response representations of linear filters, monotone treatment response and partial identification on a finite grid are established tools, and are used here as tools: of the results below, two are not applications of an existing framework, and the rest are derivations recorded for completeness. What the problem adds is their interaction: persistent deployment, a capacity estimand, interference between units and arbitrary calendar shocks appear together, and the constraints that combination imposes on any design are the subject of the paper.

The paper belongs to a literature on experimenting when the intervention moves the environment the units share. \citet{wagerxu2021} study experiments whose treatment shifts a market-clearing quantity, and \citet{munro2025equilibrium} give conditions under which unit-level contrasts recover equilibrium effects. In both the common component is a nuisance to be removed; here it is part of the object being measured, which reverses what a design should want from it.

Four further connections are worth naming. The kernel representation is the impulse-response form of a causal linear time-invariant system acting on a transformed input, whose linear part is the object of the distributed-lag tradition beginning with \citet{almon1965}. Set identification on a finite grid is partial identification in the sense of \citet{manski2003} and \citet{tamer2010}. The minimum-variance weights are the generalised least-squares solution of \citet{aitken1935}, and allocating a fixed budget across arms is the optimal design problem of \citet{kiefer1959} and \citet{pukelsheim2006}. Adaptive refinement of the capacity interval is threshold search: stochastic approximation in the sense of \citet{robbinsmonro1951}, and closer in spirit to the sequential dose-finding designs of \citet{oquigley1990}, where the target is a crossing rather than a curve. \citet{xiong2024staggered} solve for the treatment-start times that maximise power when assignment cannot be reversed, which is the cost of staggering from the opposite direction, and for simultaneous inference across arms we draw on \citet{cck2014}.

\section{Methods}

\subsection{The capacity experiment}
\label{sec:design}

 What the experiment identifies and what it costs are the subject of the two sections that follow; everything else calibrates it, validates it, or extends it to a different question.

\subsubsection{Environment}

Indices are used throughout as follows: $i$ for a signal, $p$ for a sleeve running that signal, $b$ for a block of consecutive periods, $\ell$ for a period within a block, and $t$ for the calendar date. Where a single signal is under study the index $i$ is suppressed.

Deployment is assumed to enter the conditional mean through an erosion map:
\begin{equation}
 \E[A_{i,t}(w)\mid\Fcal_t]=\mu_{i,t}-g_i(w;Z_t),
 \label{eq:erosion}
\end{equation}
with $g_i$ nonnegative, nondecreasing in each coordinate, and $g_i(\mathbf 0)=0$, and where $Z_t\subset\Fcal_t$ collects potential moderators such as liquidity, volatility, or funding conditions. The uncrowded edge $\mu_{i,t}$ is the object the allocation decision needs and the object returns cannot see.

\subsubsection{Assumptions}

The assumptions below are maintained throughout. They fix notation and the sampling scheme, restrict how sleeves interfere with one another, and state what publication is taken to identify.

\begin{assumption}[Consistency]\label{as:consistency}
Indices are $i$ for a signal, $p$ for a sleeve, $b$ for a block, $\ell$ for a period within a block and $t$ for the calendar date; $i$ is suppressed when a single signal is under study. By \Cref{as:consistency} the observed adjusted return is a function of the full deployment vector, $A_{i,t}=A_{i,t}(W_t)$.
\end{assumption}

Interference across time is handled by \Cref{as:noanticipation}. Interference across sleeves is the harder problem: a sleeve deployed at a high scale moves the prices and the borrow the others face, so the control arm is treated indirectly. We restrict how that interference enters rather than assuming it away.

\begin{assumption}[Exposure mapping]\label{as:exposure}
Let $N$ be the number of sleeves running the strategy, of which $P\leq N$ are in the experiment. Let $\Gamma=(\gamma_{pq})$ be the overlap matrix, with $\gamma_{pq}\ge0$ the share of sleeve $q$'s deployment reaching the securities sleeve $p$ holds and $\sum_q\gamma_{pq}=1$. Write $z_b=(z_{1b},\dots,z_{Nb})$ for the vector of scales assigned in block $b$ and $A_{pb\ell}(z_b)$ for the potential adjusted return of sleeve $p$ in period $\ell$ under that vector. Define the exposure of sleeve $p$ as the pair
\[
 e_p(z_b)=\bigl(z_{pb},\;\gamma_p^{\top}z_b\bigr),
\]
and assume exposure consistency: $A_{pb\ell}(z_b)=A_{pb\ell}(z_b')$ whenever $e_p(z_b)=e_p(z_b')$. Erosion then evolves as
\[
 h_{p,t+1}=a\,h_{p,t}+(1-a)\bigl\{c^{\mathrm{own}}(W_{pt})+c^{\mathrm{agg}}(\gamma_p^{\top}W_t)\bigr\},
\]
with $c^{\mathrm{own}}$ and $c^{\mathrm{agg}}$ continuous, non-decreasing, zero at zero and differentiable at $\bar W_t$.
\end{assumption}

The homogeneous case $\gamma_{pq}=N^{-1}$ reduces the argument to average deployment $\bar W_t$ and makes it common to every sleeve at a date; that is the balanced design of this section and the case in which the results below are stated. Two restrictions are being imposed. A sleeve does not care which other sleeve deployed the capital, only how much reached its own securities; and the crowding it faces scales with the average rather than the total, so that adding a sleeve at an unchanged average leaves it unaffected. The second is what makes the experiment's own footprint small: moving one sleeve from $\beta_0$ to $\beta$ moves the average by $(\beta-\beta_0)/N$, not by $\beta-\beta_0$. Neither restriction is innocuous, and both are what make a sleeve-level design interpretable at all.

\begin{assumption}[Block randomisation]\label{as:unconfounded}
Blocks are indexed by $b$ and periods within a block by $\ell=1,\dots,L$. The vector $Z_b$ of scales, drawn from the finite arm set $\{\beta_0,\dots,\beta_J\}^N$, is assigned at the start of block $b$ and held for its duration, and
\[
 Z_b\;\perp\;\bigl\{A_{pb\ell}(z)\bigr\}_{z,p,\ell}\;\Bigm|\;\Hcal_b ,
\]
where $\Hcal_b$ is the history entering block $b$, including the inherited erosion state. Randomisation is therefore over the whole vector, not over one sleeve's policy holding the others fixed, which is what \Cref{as:exposure} requires.
\end{assumption}

\begin{assumption}[Arm overlap]\label{as:overlap}
$\Prob(Z_{pb}=\beta_j\mid\Hcal_b)\geq\underline p>0$ for every arm $j$ and every block, and within each block both a treated and a control sleeve are assigned. The design assigns a fixed number of sleeves to each arm and overlap is homogeneous, so $\gamma_p^{\top}Z_b=\bar W_b$ is the same for every sleeve and every vector in the support.
\end{assumption}

\begin{remark}[Assigned and realised deployment]\label{rem:ZW}
$Z_{pb}$ is the assigned scale and $W_{pb\ell}$ the deployment realised in period $\ell$ of block $b$ by sleeve $p$. They coincide under full compliance; they do not under imperfect implementation, competitor response or measurement error, and the estimands below are defined on assignment unless realised deployment is measured, in which case $Z$ serves as an instrument for $W$.
\end{remark}

\begin{assumption}[Edge invariance]The assignment mechanism affects $A_{i,t}$ only through $W_t$: the uncrowded edge $\mu_{i,t}$, the loadings $b_{i,t}$, and the law of $\varepsilon_{i,t+1}$ do not depend on the assignment rule.
\end{assumption}

\begin{assumption}[No anticipation and bounded carryover]\label{as:noanticipation}
$A_{i,s}(w)$ does not depend on assignments dated after $s$, and the erosion state satisfies a recursion $h_{i,t+1}=\Phi_i(h_{i,t},W_t)$ that is a uniform contraction under every constant policy in the experimental support: there is $r<1$ with $|\Phi_i(h,w)-\Phi_i(h',w)|\leq r|h-h'|$ for all $h,h'$ and every such $w$. A unique steady state under each constant policy, independent of the state the block starts from, follows.
\end{assumption}

\begin{assumption}[Erosion primitives]\label{as:linear}
There is a scale response $c:[0,\bar\beta]\to[0,\infty)$ with $c(0)=0$, continuous and non-decreasing, such that the erosion state satisfies: (i) it is scalar; (ii) it responds linearly to its own lag and to the scale response of current deployment; (iii) it is first-order Markov; (iv) it is stable and time homogeneous; and (v) under constant deployment $\beta$ its steady state is $c(\beta)$.
\end{assumption}

Price impact is concave in size \citep{toth2011,sato2024}, whereas the working scale response used below is convex. The two are consistent because $c$ is broader than instantaneous execution cost: it aggregates the impact of a whole holding period, the leakage of a persistent position into the borrow and the response of competitors, and those channels can rise more than proportionally with deployment even where the instantaneous impact of a single order does not. Since a linear map calibrated at a reference scale understates erosion below it and overstates it above, the results hold to the accuracy of a local linearisation over the range of $\beta$ the design visits.

\Cref{as:unconfounded} is automatic under randomisation and is what observational studies of deployment and returns cannot supply, since a manager who scales up after a good run makes assignment a function of the outcome.

Regressing realised net returns on realised turnover estimates a mixture of the capacity slope and the manager's own scaling rule, which is why the estimand is defined on assignment.

\subsubsection{The estimands, formally}

Deployment is not a scalar: the economically relevant treatment is the policy of \Cref{def:policy}, a history of levels, and the estimands below are therefore indexed by a policy and by the length of time it is held.

\begin{definition}[Constant-scale policies]\label{def:policy}
For a scale $\beta$ and a hold length $L$, write $\bar W(\beta,L)=(\beta\pi,\dots,\beta\pi)$ for the policy that holds scale $\beta$ for $L$ consecutive periods, and $\bar W(0,L)$ for the corresponding withdrawal policy. All estimands below are indexed by a policy and by the erosion state $h_{i,t_0}$ inherited at the start of the block, and all comparisons hold that inherited state and the pre-block history fixed.
\end{definition}

\begin{definition}[Capacity curves and capacity]\label{def:estimands}
Fix an allocation direction $\pi$ with $\norm{\pi}=1$ and a hold length $L$; \Cref{def:estimands} names the objects the design targets. The finite-hold curves are
\[
 m^{\mathrm{end}}_{i,L}(\beta)=\E\!\left[A_{i,t_0+L}\bigl(\bar W(\beta,L);h_{i,t_0}\bigr)\right],
 \qquad
 m^{\mathrm{avg}}_{i,L}(\beta)=\E\!\left[\frac{1}{L}\sum_{\ell=1}^{L}A_{i,t_0+\ell}\bigl(\bar W(\beta,L);h_{i,t_0}\bigr)\right],
\]
the steady-state curve is $m_i^{\infty}=\lim_{L\to\infty}m^{\mathrm{end}}_{i,L}=\lim_{L\to\infty}m^{\mathrm{avg}}_{i,L}$, and marginal crowding is $\kappa_i(\beta)=c_i'(\beta)$. Writing $\bar\beta$ for the mandate ceiling and $\tau_{\mathrm{out}}$ for the withdrawal hurdle,
\[
 \bown[,i](\bar w)=\sup\bigl\{\beta\in[0,\bar\beta]:m_i^{\infty}(\beta;\bar w)\geq\tau_{\mathrm{out}}\bigr\},
 \qquad
 \btot[,i]=\sup\bigl\{\beta:\mu_i-c^{\mathrm{own}}_i(\beta)-c^{\mathrm{agg}}_i(\beta)\geq\tau_{\mathrm{out}}\bigr\}.
\]
\end{definition}

Three remarks follow. The cost calculation uses $m^{\mathrm{avg}}$, which employs every observation, and $m^{\mathrm{end}}$ serves as the benchmark for a design that discards within-block information. Under \cref{eq:stock} the steady-state erosion at constant deployment $\beta$ is $c_i(\beta)$, so the capacity curve inherits whatever curvature the scale response has while the kernel governs only the speed of accumulation. And by \Cref{prop:sharp} a design with finitely many arms identifies the curve at the arms and nowhere else, so what it delivers for capacity is an interval rather than a point; what a return history alone identifies, and therefore what the experiment has to improve on, is recorded in \Cref{app:boundary}.

\begin{proposition}[The sharp identified set for capacity]\label{prop:sharp}
Let arms $\beta_0<\dots<\beta_J$ be assigned and let $m$ be continuous and non-increasing with $m(\beta_j)\geq\tau_{\mathrm{out}}>m(\beta_{j+1})$. Capacity is a level crossing of a monotone response, so what a finite grid delivers is the sharp set of \citet{manski1997} specialised to a crossing. The sharp identified set for $\bown(\bar w)$, the own-sleeve capacity at the average the design holds fixed, is
\[
 \bigl(\beta_j,\beta_{j+1}\bigr)\ \text{ if } m(\beta_j)>\tau_{\mathrm{out}},
 \qquad
 \bigl[\beta_j,\beta_{j+1}\bigr)\ \text{ if } m(\beta_j)=\tau_{\mathrm{out}} .
\]
\end{proposition}

The closed interval $[\beta_j,\beta_{j+1}]$ is a valid outer bound but is not sharp: continuity excludes $\beta_{j+1}$ because the curve is already below the hurdle there, and excludes $\beta_j$ when the curve is strictly above it, since it then remains above on a neighbourhood to the right. A point estimate requires a parametric form, an interpolation rule, or a curvature restriction.

Sharpness concerns the identified set, not the set a design reports, and the distinction is not a fine one: at the design size the interval between two estimated arm means covers the truth less than half the time, and refining the grid sequentially drives that below a fifth. A reporting rule with coverage has to be constructed separately.

\begin{proposition}[A capacity set with coverage]\label{prop:band}
Let arms $\beta_1<\dots<\beta_J$ be assigned, possibly in stages, and let $[\widehat m(\beta_j)\pm c\,\widehat{\mathrm{se}}_j]$ be a band that covers every arm mean simultaneously with probability at least $1-\alpha$. Under monotonicity of $m$, report
\[
 \widehat B=\bigl[\max\{\beta_j:\widehat m(\beta_j)-c\,\widehat{\mathrm{se}}_j\geq\tau_{\mathrm{out}}\},\;
 \min\{\beta_j:\widehat m(\beta_j)+c\,\widehat{\mathrm{se}}_j<\tau_{\mathrm{out}}\}\bigr],
\]
with the endpoints set to $0$ and $\bar\beta$ when the corresponding sets are empty. Reporting a set for a partially identified parameter is the problem of \citet{imbensmanski2004}, refined by \citet{stoye2009}; the construction below is the test-inversion version adapted to a design whose arms are chosen sequentially. Suppose that arms are drawn from a candidate set $\mathcal G$ fixed before the experiment, with $|\mathcal G|=M$ finite, and that the band is simultaneous over all of $\mathcal G$. Then $\Prob(\bown\in\widehat B)\geq1-\alpha$ whatever the placement rule and however many stages were used.
\end{proposition}

The prespecified candidate set is what makes the statement hold under adaptive selection. With a data-dependent rule the realised arms are random, so a band calibrated on them alone is not valid, and controlling the number of selected points does not help if the locations are drawn from a continuum. Fixing $\mathcal G$ in advance and correcting over it removes the selection entirely, at the cost of a multiplier growing in $M$. Stagewise splitting with conditionally valid intervals, alpha spending across stages, confidence sequences, or a process-level band over a continuum would each serve instead and would be less conservative; the implementation below uses the simplest, a Bonferroni correction over the whole of $\mathcal G$ rather than over the arms a rule happens to select, and the resulting coverage of essentially one against a nominal $0.90$ shows how much is given away.

 With a data-dependent rule the realised arms are random, so a band calibrated only on them is not in general valid; the exercises below use a Bonferroni adjustment over the maximum number of arms the rule can produce, which restores validity conservatively.

Unless stated otherwise $m_i$ denotes $m_i^{\infty}$. The distinction is not cosmetic: the sampling rule above shows that the same nominal scale identifies $m^{\mathrm{avg}}_{i,L}$ rather than $m_i^{\infty}$, and that the gap between them is first order at empirically relevant persistence.

\subsection{Why the obvious design answers the wrong question}
\label{sec:atten}

A design that assigns a scale and holds it for a fixed number of periods does not measure the steady-state response, because the erosion stock has not finished accumulating when the block ends. How much is missing depends on how the block is summarised, so the choice of outcome is a choice of estimand rather than a detail of estimation.

Write $F_j=\sum_{s<j}\omega_s$ for the fraction of the steady-state effect accumulated by the $j$th period of a block. A block that holds scale $\beta$ for $L$ periods and reports the convex average $\sum_ju_jA_{t_0+j}$ then identifies $-c(\beta)R_L(u)$ with $R_L(u)=\sum_ju_jF_j$, since deployment is $\beta$ in the $j$ periods elapsed and zero before. Sampling the last observation gives $R_L=F_L$ and the block average gives $R_L=G_L\equiv L^{-1}\sum_{j\le L}F_j$; under \cref{eq:stock}, $F_L=1-a^{L}$ and $G_L=1-a(1-a^{L})/\{L(1-a)\}$. Monotonicity of $F$ gives $G_L\leq F_L$, so averaging attenuates at least as much as sampling the end, and if the kernel has mass beyond every finite horizon then $F_j<1$ for all finite $j$ and no convex sampling rule recovers the steady state in finite time; a kernel of finite memory $S$ has $F_L=1$ for $L>S$ and does. Under \Cref{eq:stock}, $F_j=1-a^{j}$.

Because outcomes are indexed by exposures rather than by own assignments, three effects can be asked of such an experiment and they differ. Write $\Delta=\beta-\beta_0$ for the contrast in assigned scales. The controlled direct effect is the response of a sleeve to its own scale holding average deployment fixed,
\[
 \mathrm{CDE}(\beta,\beta_0)=-\bigl\{c^{\mathrm{own}}(\beta)-c^{\mathrm{own}}(\beta_0)\bigr\},
\]
the total effect of the sleeve's assignment adds the crowding the sleeve induces on itself by moving the average, and the strategy-level effect is what happens when every sleeve moves together, so that the average moves by $\Delta$ rather than by $\Delta/N$:
\[
 \mathrm{TE}^{\mathrm{strat}}(\beta,\beta_0)=\mathrm{CDE}(\beta,\beta_0)-\bigl\{c^{\mathrm{agg}}(\bar W+\Delta)-c^{\mathrm{agg}}(\bar W)\bigr\}.
\]
It is the last of these that a capacity statement about a strategy refers to.

\begin{proposition}[What a sleeve-level contrast identifies]\label{prop:own}
Under \Cref{as:exposure} and \Cref{as:overlap}, let two sleeves be assigned $\beta$ and $\beta_0$ within the same block. Because the design fixes $\bar W_b$ across the support, the two exposures share their second coordinate, and the within-date arm contrast identifies
$\mathrm{CDE}(\beta,\beta_0)R_L(u)$ exactly. It does not identify $\mathrm{TE}^{\mathrm{strat}}$, which differs from $\mathrm{CDE}$ by a term of order one.
\end{proposition}

\begin{remark}[Intervening on one sleeve is a third thing]\label{rem:onesleeve}
The controlled direct effect is not the effect of intervening on a single sleeve's scale while holding every other sleeve fixed. That intervention moves $\bar W$ by $\Delta/N$, so by differentiability of $c^{\mathrm{agg}}$ it differs from $\mathrm{CDE}$ by $c^{\mathrm{agg}\prime}(\bar W)\Delta/N+o(N^{-1})$ and coincides with it only in the limit of many sleeves. The design delivers the first, not the second; the distinction matters when $N$ is small enough that one sleeve is a material share of the strategy.
\end{remark}

\begin{remark}[What this design does not identify]\label{rem:agg}
The gap between what the design delivers and what a capacity statement about a strategy asks for is not a matter of precision but of support, and by \Cref{rem:onesleeve} it is not the gap between the contrast and a one-sleeve intervention either: the balanced assignment that keeps the calendar component out of the contrast also keeps average deployment constant across every vector the design can produce, so the second coordinate of the exposure never varies. The contrast holds average deployment essentially fixed and varies one sleeve within it; a capacity statement varies the average itself. The two coincide only if $c^{\mathrm{agg}}$ is flat, that is if crowding is generated by a sleeve's own execution rather than by the accumulated position of everyone running the signal, which is the opposite of what the crowding literature reports. Where the second channel matters, the design below understates capacity erosion, and identifying the total effect requires variation in $\bar W$ itself: randomising deployment across non-overlapping security partitions or markets, with sleeves used only for measurement, or randomising at the book level across periods and accepting the calendar exposure that \Cref{prop:rank} shows to be costly. This is the price of contemporaneous assignment, and \Cref{prop:tradeoff} shows it is not an accident of the design but a property of any within-date contrast.
\end{remark}

Under \Cref{eq:stock}, $F_j=1-a^{j}$, so $F_L=1-a^{L}$ and $G_L=1-a(1-a^{L})/\{L(1-a)\}$. Since $F_j$ is non-decreasing, $G_L\leq F_L$: averaging attenuates at least as much as sampling the end. If the accumulation kernel has positive mass beyond every finite horizon, then $F_j<1$ for every finite $j$ and no uncorrected convex sampling functional recovers the steady-state effect in finite time; a kernel of finite memory $S$ has $F_L=1$ for $L>S$ and does.

\Cref{prop:own} fixes what the contrast measures; the distinction between the two families of weights that follow fixes how it is measured. Convex weights $u$ define which finite-hold estimand is targeted, and every one of them is attenuated. Weights $v$ satisfying $v^{\top}F=1$ define instead a deattenuated estimator of the steady-state effect: they are not a sampling rule, they use knowledge of $F$, and they trade the model-free character of $\Delta_L(u)$ for an unbiased recovery of $\Delta_\infty$.

\begin{proposition}[Oracle minimum-variance recovery of the steady-state effect]\label{prop:optw}
Let $D=(D_1,\dots,D_L)^{\top}$ collect the within-period arm contrasts of a block, so that $\E[D]=F\,\Delta_\infty$ with $F=(F_1,\dots,F_L)^{\top}$ and $\Delta_\infty=-c(\beta)$, and let $\Var(D)=\Sigma$ be positive definite. Among linear estimators $\widehat\Delta_\infty=v^{\top}D$ that are unbiased for $\Delta_\infty$, that is those with $v^{\top}F=1$, the variance is minimised uniquely at
\begin{equation}
 v^{*}=\frac{\Sigma^{-1}F}{F^{\top}\Sigma^{-1}F},
 \qquad \Var(v^{*\top}D)=\frac{1}{F^{\top}\Sigma^{-1}F}.
 \label{eq:optw}
\end{equation}
Under $\Sigma=\sigma^{2}I$ this is $v^{*}=F/(F^{\top}F)$. 
\end{proposition}

Under $\Sigma=\sigma^{2}I$ the weights are proportional to $F$ and therefore increase with the period index, so early observations are downweighted rather than discarded. That monotonicity is a property of the scalar case and does not extend: for a general positive-definite $\Sigma$ the vector $\Sigma^{-1}F$ need be neither monotone nor nonnegative, and periods whose errors are strongly correlated with the rest can receive negative weight. Unbiasedness is in any case conditional on $F$ being correct and known, and the variance ignores the cost of learning $F$ and $\Sigma$.

\Cref{tab:weights} evaluates the four at the calibrated persistence, and three things follow. Sampling only the terminal observation is inefficient by a factor of three against the block average, which is the cost of discarding the path. The oracle improves on the block average by seven per cent, because at a geometric kernel the optimal weights are close to uniform once the early periods have been downweighted: the gain from optimal weighting is real and small. And the feasible estimator, which learns the kernel from the trajectory it deattenuates, is dominated at every sample size considered: its bias at design size is four times the target and its standard deviation fourteen times the block average's, and it remains the worst of the four even a hundredfold larger. Deattenuation is worth doing with a transported kernel and not worth doing with an estimated one.

\begin{table}[t]
\centering
\caption{Deattenuated recovery of the steady-state effect}
\label{tab:weights}
\small
\begin{tabular}{@{}lrrrrrrrr@{}}
\toprule
& \multicolumn{2}{c}{Terminal} & \multicolumn{2}{c}{Block average} & \multicolumn{2}{c}{Oracle} & \multicolumn{2}{c}{Feasible} \\
\cmidrule(lr){2-3}\cmidrule(lr){4-5}\cmidrule(lr){6-7}\cmidrule(lr){8-9}
Sample & bias & s.d. & bias & s.d. & bias & s.d. & bias & s.d. \\
\midrule
Design size & $-0.000$ & 0.131 & $-0.000$ & 0.039 & $-0.000$ & 0.037 & $-0.537$ & 0.540 \\
Ten times larger & $-0.000$ & 0.021 & $+0.000$ & 0.006 & $-0.000$ & 0.006 & $-0.052$ & 0.182 \\
Hundred times & $-0.000$ & 0.007 & $+0.000$ & 0.002 & $+0.000$ & 0.002 & $-0.002$ & 0.013 \\
\midrule
Closed-form s.d., design size & & 0.131 & & 0.039 & & 0.037 & & --- \\
\bottomrule
\end{tabular}

\vspace{0.35em}
\begin{minipage}{0.96\textwidth}\footnotesize
20{,}000 replications at design size and 4{,}000 and 2{,}000 at the larger scales, two-year holds, geometric kernel at $a=0.9177$, truth $-0.123$. Design size is a hundred sleeves split evenly between the two arms and forty blocks. The first three estimators use the transported $F$ and, for the oracle, the true $\Sigma$; the feasible estimator fits the kernel to the observed trajectory, which \Cref{prop:traj} shows is identified up to scale. The last row gives the closed-form standard deviations $\mathrm{se}/\{F_L\sqrt{n}\}$, $\sqrt{L\,\mathrm{se}^{2}/n}/(LG_L)$ and $(F^{\top}\Sigma^{-1}F)^{-1/2}$, which the simulation reproduces to the third decimal.
\end{minipage}
\end{table}

This is an estimand failure rather than an inference failure: period-by-period randomisation delivers valid inference for the impact-period effect and becomes a failed capacity design only when that effect is read as the steady-state slope. Short holds therefore succeed for the impact-period effect and fail for steady-state capacity, and end-of-block and block-average sampling remain different estimands rather than competing estimators of one object.

\subsubsection{What the trajectory identifies, and the robust answer}

The convention by which the outcome is adjusted is a further estimand choice, developed in \Cref{app:outcome}. A design that records only the block outcome also discards the path. Recording the whole path costs nothing and identifies more.

\begin{proposition}[Trajectory identification and the one-sided bound]\label{prop:traj}
Let the within-block arm contrasts $D_1,\dots,D_L$ be observed at a single scale $\beta$, so that $\E[D_j]=-c(\beta) F_j$. Then, under \Cref{prop:kernel} and without any parametric restriction on the kernel:
\begin{enumerate}[label=(\roman*),leftmargin=*]
\item the shape $F_j/F_L$ is identified for $j\le L$, and hence the increments $\omega_0/F_L,\dots,\omega_{L-1}/F_L$;
\item since $F_L\le1$, the terminal contrast bounds the steady-state effect from below, $|\Delta_\infty|\ge|\E[D_L]|$, with no kernel assumption;
\item $\Delta_\infty$ is point identified only if $F_L$ is bounded below, a restriction on the unobserved tail that the design cannot supply.
\end{enumerate} 
\end{proposition}

The proposition splits the empirical content of the design into two answers that should be reported together. The robust answer is the terminal contrast: it is model free, it bounds the steady-state effect from one side, and at the calibrated persistence a two-year hold leaves a shortfall of thirteen per cent and a four-year hold of under two. The model-assisted answer deattenuates by an assumed $F$, which is what \Cref{prop:optw} requires and what the calibration below supplies; it is a point estimate, and its error is multiplicative in the kernel error rather than additive. A design that reports only the second hides where the conclusion comes from.

\subsection{How replication works}

Replication is an identification question before it is a variance question, and \Cref{app:support} separates the three parts. The full vector of arm effects is identified if and only if the matrix of arm indicators has full rank after projecting out sleeve and date effects, which fails when every sleeve receives the same scale on a date, and a particular contrast is identified if and only if it lies in the row space of that residualised matrix; a contrast is free of an arbitrary calendar component if and only if its weights sum to zero; and among such contrasts the efficient one solves a constrained least-squares problem, of which the equicorrelated formula below is the symmetric case. Each is standard, and it is their conjunction that constrains the design.

The same setting settles a question the design cannot avoid, and the answer is the sharpest constraint in the paper.

\begin{proposition}[When aggregate crowding is identified within a date]\label{prop:tradeoff}
Let the adjusted return of sleeve $p$ at date $t$ satisfy
\[
 Y_{pt}=\mu_t-c^{\mathrm{own}}(W_{pt})-c^{\mathrm{agg}}(\gamma_p^{\top}W_t)+\varepsilon_{pt},
\]
with the calendar component $\mu_t$ unrestricted, and write $g_t=\bigl(c^{\mathrm{agg}}(\gamma_1^{\top}W_t),\dots,c^{\mathrm{agg}}(\gamma_P^{\top}W_t)\bigr)^{\top}$.
\begin{enumerate}[label=(\roman*),leftmargin=*]
\item If every row of $\Gamma$ equals a common vector $\gamma^{\top}$, then $g_t=c^{\mathrm{agg}}(\gamma^{\top}W_t)\iota\in\operatorname{span}(\iota)$ and $(\mu_t,c^{\mathrm{agg}})$ and $(\mu_t-c^{\mathrm{agg}}(\gamma^{\top}W_t),0)$ generate the same joint law of $\{Y_{pt}\}_{p\le P}$ at every date: aggregate crowding is not identified from within-date variation by any estimator. Uniform overlap, $\Gamma=N^{-1}\iota\iota^{\top}$, is the special case $\gamma^{\top}W_t=\bar W_t$.
\item If $g_t\notin\operatorname{span}(\iota)$, there exist contrasts with $q^{\top}\iota=0$ and $q^{\top}g_t\neq0$, so robustness to arbitrary calendar effects and information about aggregate crowding are attainable together.
\end{enumerate}
\end{proposition}

\begin{corollary}[Zero-sum contrasts under homogeneous overlap]\label{cor:zerosum}
Under (i), a linear contrast eliminates $\mu_t$ for arbitrary $\mu_t$ if and only if $q^{\top}\iota=0$, and since $g_t\in\operatorname{span}(\iota)$ any such contrast eliminates $g_t$ as well. Robustness and information about $c^{\mathrm{agg}}$ are then mutually exclusive among linear contrasts, which is the identification failure of (i) seen through one class of estimators.
It is proved with \Cref{prop:tradeoff}.
\end{corollary}

\begin{corollary}[Tracing aggregate crowding under heterogeneous overlap]\label{cor:trace}
Suppose two sleeves receive the same own-treatment arm but have aggregate-exposure loadings $\gamma_p^{\top}\neq\gamma_q^{\top}$ that are fixed over the experiment and known up to a common positive scale, or consistently estimable from data independent of assignment and of the experimental outcome shocks. Their within-date difference eliminates the unrestricted calendar component and the own-sleeve response, and retains $c^{\mathrm{agg}}(\gamma_p^{\top}W_t)-c^{\mathrm{agg}}(\gamma_q^{\top}W_t)$. Variation in aggregate deployment across dates then traces $c^{\mathrm{agg}}$ over the support the design visits, up to the normalisation of the loadings.
\end{corollary}

\begin{remark}[Overlap as a design instrument]\label{rem:overlap}
The impossibility is a property of homogeneous overlap, not of within-date contrasts as such. A design that deliberately induces heterogeneity in $\Gamma$, by running sleeves on disjoint regional sub-universes, separate borrow pools or distinct liquidity tiers, moves $g_t$ out of $\operatorname{span}(\iota)$ and recovers the second estimand, at the price of a weaker exposure mapping and of having to estimate $\Gamma$. What governs the gain is $\lVert M_\iota g_t\rVert$, the component of aggregate crowding orthogonal to the calendar direction, which the balanced design of this section sets to zero by construction. At the calibration used below, six sleeves split into two disjoint blocks give $\lVert M_\iota g_t\rVert=0.27$ against $0$ under homogeneous overlap.
\end{remark}

The constructive reading of \Cref{prop:tradeoff} is the more useful one. Identifying $c^{\mathrm{agg}}$ requires variation in average deployment, which is variation across dates, and any estimator using it is exposed to the calendar component the zero-sum condition removes; the price is the staggered inflation quantified below, a factor of $23.9$ at a hundred sleeves. The choice is between two estimands at very different prices, and contemporaneous assignment selects the own-sleeve response at low cost over the aggregate one at high cost.

The sample a design requires follows from the achieved gap of \Cref{sec:atten}, the residual volatility and the dependence between sleeves.

Write $\widehat D_t$ for the within-period arm contrast, $\Omega_D$ for its long-run variance, $g$ for the steady-state gap between the arms, $T$ for the number of calendar periods, and $c_{\alpha,\pi}=z_{1-\alpha}-z_{1-\pi}$ for the normal constant delivering size $\alpha$ against power $\pi$.

\begin{corollary}[Required scale under stock attenuation and correlated sleeves]\label{prop:feasibility}
Let $P$ sleeves be split evenly between a treated and a control arm within the same calendar periods and let assigned policies be held for $L$ periods with the block average as the outcome.  Then $T=c_{\alpha,\pi}^{2}\,\Omega_D/[G_L^{2}g^{2}]$. If in addition the period residuals have common variance $\sigma^2$, average pairwise correlation $\bar\rho$, and no serial dependence within or across blocks, then
\[
 \Var\bigl(\widehat D_t\bigr)=\frac{4\sigma^{2}\,(1-\bar\rho)}{P},
\]
so that, in that special case, detecting an achieved gap $G_Lg$ with size $\alpha$ and power $\pi$ requires
\[
 T\;=\;\frac{4\,\sigma^{2}c_{\alpha,\pi}^{2}\,(1-\bar\rho)}{P\,G_L^{2}\,g^{2}},
 \qquad G_L=1-\frac{a(1-a^{L})}{L(1-a)} .
\]
\end{corollary}

Under staggered assignment the arm contrast is formed across rather than within periods, the common component no longer cancels, and the variance of an equicorrelated mean gives instead
\[
 T^{\mathrm{stag}}\;=\;\frac{4\,\sigma^{2}c_{\alpha,\pi}^{2}\,\bigl[1+(P-1)\bar\rho\bigr]}{P\,G_L^{2}\,g^{2}},
\]
which converges to a strictly positive limit as $P\to\infty$: replication saturates.

\begin{remark}[Order of the approximation]\label{rem:firstorder}
The requirement is first order in the normal approximation to the sampling distribution of the arm contrast. It is not exact in small samples, under randomisation inference, under sequential stopping or adaptive allocation, or when $\Omega_D$ is itself uncertain.
\end{remark}

\begin{remark}[What the sleeves are]\label{rem:sleeves}
The expression treats the $P$ sleeves as carrying a common effect. That is immediate when they are parallel implementations of one signal in different books or execution channels. When they are distinct strategies, as in the panel used for calibration, the estimand is an average capacity response across them, and the expression prices a design targeting that average rather than any individual capacity.
\end{remark}

The remaining proofs are collected in \Cref{app:proofs}.

\subsection{The cost of an experiment and the bound an impact model gives}

Two further results price the design and place it against the alternative. The first says that arms cannot be made cheap by making them narrow; the second that the impact models desks already run bound capacity from one side only. A third fixes what a grid can resolve.

\begin{proposition}[The local opportunity cost of an experiment]\label{prop:straddle}
Let $V$ be three times differentiable with an interior maximiser $\beta^{*}$, so that $V'(\beta^{*})=0$ and $V''(\beta^{*})<0$, and let $c$ be differentiable and strictly increasing. For arms placed symmetrically at $\beta^{*}\pm d$, a Taylor expansion gives $\Lambda=-\tfrac12V''(\beta^{*})d^{2}+o(d^{2})$ and $c(\beta^{*}+d)-c(\beta^{*}-d)=2c'(\beta^{*})d+o(d)$, so the total forgone edge converges as $d\to0$ to
\[
 C^{*}=\frac{K\,\{-V''(\beta^{*})\}}{8\,\{c'(\beta^{*})\}^{2}},
 \qquad K=\frac{c_{\alpha,\pi}^{2}\Omega_D}{G_L^{2}},
\]
which is strictly positive. Narrowing the arms around the operating scale therefore does not make the experiment cheap: the second-order saving in forgone edge is exactly offset by the squared first-order loss in the identifying gap. Under the working response $c(\beta)=\kappa\beta+\zeta\beta^{2}$ the cancellation is exact and $C^{*}$ is attained at every $d$.
\end{proposition}

\begin{proposition}[What an impact model identifies]\label{prop:impact}
Suppose total erosion decomposes into an execution channel and a crowding channel,
\[
 c(\beta)=c^{\mathrm{exec}}(\beta)+c^{\mathrm{crowd}}(\beta),
\]
both non-negative, increasing and zero at zero, where $c^{\mathrm{exec}}$ is the expected round-trip cost of trading the position at scale $\beta$ and $c^{\mathrm{crowd}}$ is the loss of gross edge caused by accumulated positioning in the signal. Assume the impact model is correctly specified for $c^{\mathrm{exec}}$ and identified from the execution record, and that its estimand is a cost of trading that absorbs no part of the decay in gross edge. Then it identifies $c^{\mathrm{exec}}$ and is uninformative about $c^{\mathrm{crowd}}$, because execution records contain the price paid to trade and not the alpha that would have been earned had nobody else been positioned. Writing $\beta^{\mathrm{imp}}$ for the scale solving $\mu-c^{\mathrm{exec}}(\beta)=\tau_{\mathrm{out}}$,
$\beta^{\mathrm{imp}}\geq\bown$, with equality if and only if $c^{\mathrm{crowd}}$ vanishes on $[0,\bown]$. An impact model therefore bounds capacity from one side only, and the wedge is the object the experiment is built to measure.
\end{proposition}

\begin{remark}[A heuristic resolution scale]\label{rem:resolution}
Two considerations govern how short the set can be. The reported endpoints are assigned arms, so on a grid of spacing $\Delta$ the set is never shorter than $\Delta$ unless it is empty. A second and weaker consideration is the scale on which the design can separate arms at all. Let $m$ be differentiable at the crossing with $m'<0$ and let the band have half-width $c\,\mathrm{se}$. Under the local approximation $m(\beta)-\tau_{\mathrm{out}}\approx m'(\bown(\bar w))(\beta-\bown(\bar w))$, an arm is excluded only if its distance from the crossing exceeds $c\,\mathrm{se}/|m'|$ up to its own estimation error, so
$r=\mathrm{se}/|m'(\bown(\bar w))|$ is the scale on which the design can separate scales at all, and reported lengths are naturally measured in multiples of it. This is a calculation rather than a theorem: it is not a lower bound, since a favourable draw can exclude an arm whose true curve lies within the band, and a genuine local minimax statement would require a two-point testing argument over a class of curves, which we do not attempt. Reported lengths are given in multiples of $r$ throughout, which is the scale on which the comparison between placement rules is meaningful.
\end{remark}

\subsection{A separate experiment for path dependence}

Whether erosion depends only on current capital or on its history is a different question from how much capital erodes the edge, and it requires a different design. The ramp visits a sequence of levels upward and then downward, and the level-by-level differences between the two legs are zero when erosion is memoryless. The complete null is that the vector of those differences is zero; any scalar statistic is a projection of it, and no projection preserves every direction of departure, so the class of alternatives has to be declared before the statistic is chosen. Because the loop is closed by the same long holds that identify the capacity slope, the two questions require separate blocks. \Cref{app:hystprop} gives the statement, the three tests and their power.

\section{Calibration and design evaluation}

\subsection{Calibration}
\label{sec:calib}

The data price the experiment; they do not replace it: they supply the magnitudes the design results are stated in, and support the Monte Carlo exercises that validate them.

\subsubsection{Data}

The outcome is the factor-adjusted return of a long--short strategy, and we construct thirteen of them from six Fama and French portfolio sets: momentum and size spreads formed on the five-by-five size-by-prior grids for North America, Europe, Asia Pacific ex Japan and the Developed universe, a value spread from the book-to-market sorts, an investment spread from the investment sorts, and profitability and investment spreads from the five-by-five profitability-by-investment grid, together with an industry momentum spread. Returns are monthly and the common sample is 2001:01 to 2023:07, which is 271 months.

The adjustment set is selected by the commonality criterion of \citet{rodriguez2023}: from one hundred and twenty-five candidate series covering equity indices, credit, commodities and rates we retain the eight that appear most often among the strongest correlates across strategies, subject to a mutual correlation ceiling. The residual of a strategy return on these eight is the adjusted outcome, and the loadings embed the strategies in the sensitivity space whose metric describes dependence between them.

The design results depend on how fast the erosion stock dissipates, so a series is needed that measures pressure on the short side at daily frequency over a long sample. We reconstruct the cross-sectional distribution of implied borrow rates from option-implied financing spreads over 1996--2024, aggregate to monthly percentiles, and validate the reconstruction against an independently supplied monthly aggregate of the same source; the largest percentile discrepancy is below two hundredths of a percentage point.

Four quantities enter the design results. Two are read off the strategy panel; the other two are not estimated on the object they stand for but transported from a related series under an explicit bridge. The strategy-level erosion stock and the aggregate short-side series are taken to share a dissipation rate, $a_{\mathrm{strategy}}=a_{\mathrm{borrow}}$; and publication is taken to affect the adjusted return only by moving deployment, so that the post-publication decline net of a pseudo-discovery benchmark equals the total structural erosion at the deployed scale, normalised to unity, $c(1)-c(0)=0.123$.

Neither bridge is testable here, and it is worth being concrete about why. Identifying the dissipation rate from the panel itself, by projecting driver-adjusted returns on innovations in the crowding state out to twelve months and fitting a common geometric decay, returns $\widehat q=0.28$ for the four non-overlapping momentum strategies and $0.27$ for all thirteen; a twelve-month moving-block bootstrap puts the interval at $[0.00,0.99]$. It contains the transported $0.9177$, so this is no evidence against the bridge, and an exercise essentially uninformative about persistence in this sample. The second bridge fares similarly: placebo-corrected declines around publication dates run from $-0.10$ percentage points per month to $+0.58$ against a transported $c(1)=0.123$, a wide and heterogeneous distribution rather than a common constant.

The post-publication decline of published predictors, net of a pseudo-discovery benchmark, is $0.123$ percentage points per month. It measures the erosion reached at the scale deployment had attained, not a derivative at zero, so it calibrates the scale response at that scale, which we normalise to $\beta=1$: $c(1)=0.123$. Under the working curvature $\zeta=0.05$ this implies $\kappa=c'(0)=0.073$ and a marginal crowding at the reference scale of $c'(1)=\kappa+2\zeta=0.173$. Throughout, the effect the design is powered on is the contrast between the unit arm and the zero arm,
$g=c(1)-c(0)=0.123$, which is the same quantity the recovery exercises target. Calendar requirements scale with $g^{-2}$, so this definition has to be held fixed, and it is.

The exit hurdle and the curvature are design choices rather than estimates, and are stated as such in the table.

\begin{figure}[t]
\centering
\includegraphics[width=.95\textwidth]{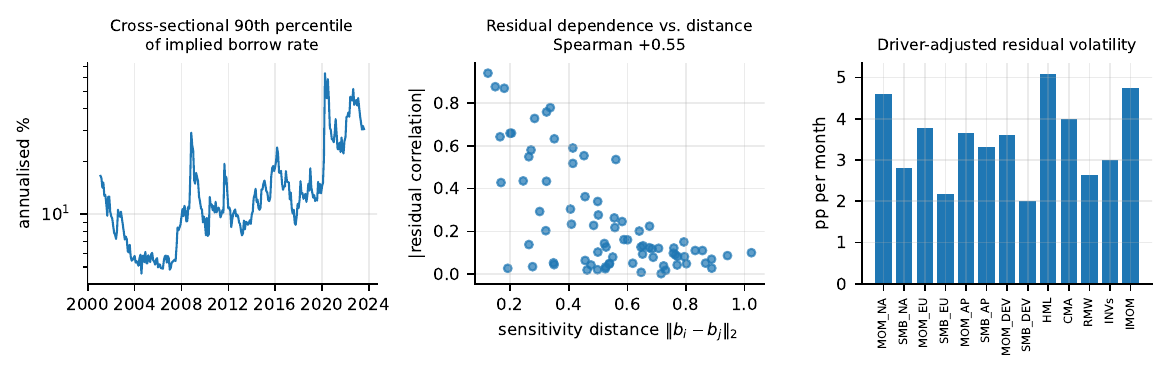}
\caption{Calibration inputs. Left: the reconstructed cross-sectional 90th percentile of implied borrow rates. Centre: absolute residual correlation between strategy pairs against distance in driver-sensitivity space. Right: driver-adjusted residual volatility by strategy.}
\label{fig:calib}
\end{figure}

\Cref{tab:calib} collects the resulting values and \Cref{fig:calib} shows the series behind them.

\begin{table}[t]
\centering
\caption{Calibration: empirically anchored scenario}
\label{tab:calib}
\small
\begin{tabular}{@{}llr@{}}
\toprule
Parameter & Source & Value \\
\midrule
$\mu$ & In-sample edge of published predictors & 0.703 \\
$\tau_{\mathrm{out}}$ & Convention & 0.150 \\
$c(1)$ & Post-publication decline, net of pseudo-discovery & 0.123 \\
$\kappa=c'(0)$ & Implied by $c(1)$ and $\zeta$ & 0.073 \\
$\sigma$ & Driver-adjusted residuals, median of 13 strategies & 3.611 \\
$\bar\rho$ & Mean pairwise correlation of those residuals & 0.186 \\
$a$ & AR(1) of detrended log borrow tail, 337 months & 0.9177 \\
$\lambda^{\top}\varphi$ & Drift of rolling loadings on the crowding state & 0.0348 \\
$\bown(\bar w)$ & Solves $\mu-c^{\mathrm{own}}(\beta)-c^{\mathrm{agg}}(\bar w)=\tau_{\mathrm{out}}$ & 2.675 \\
\bottomrule
\end{tabular}

\vspace{0.35em}
\begin{minipage}{0.94\textwidth}\footnotesize
Rates and edges are percentage points per month. The scale response used for simulation is $c(\beta)=\kappa\beta+\zeta\beta^{2}$, a member of the family of \Cref{as:linear} with increasing marginal crowding. Its curvature $\zeta$ and the implied own-sleeve capacity fix a plausible design range; neither is identified by any result in the paper, which delivers an interval rather than a point, and neither is the strategy capacity, which this design does not reach. \end{minipage}
\end{table}

\subsection{Validation: attenuation}

Each design result is verified by a Monte Carlo exercise on data generated from the model of \Cref{sec:design} at the values of \Cref{tab:calib}. The generating process is the stock recursion with the calibrated persistence and scale response, adjusted returns with the estimated residual volatility, and driver premia and loadings taken from the panel. \Cref{app:sims} records the grids, the algorithms and the replication counts in full.

\begin{figure}[t]
\centering
\includegraphics[width=.78\textwidth]{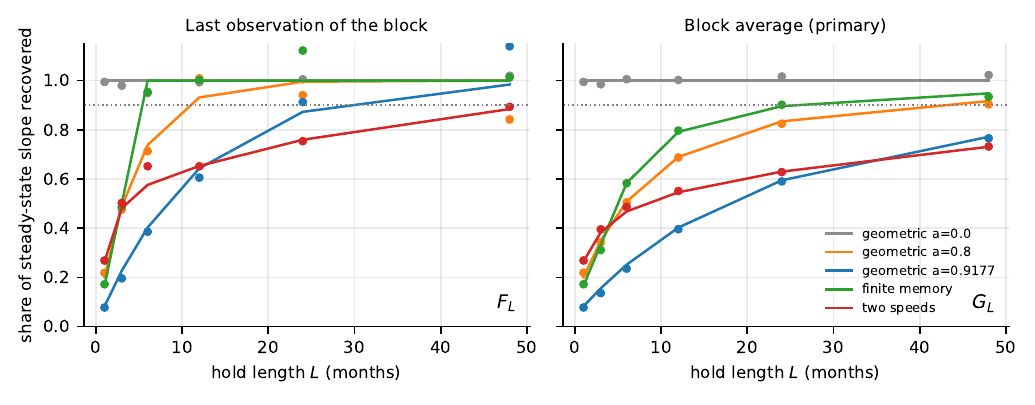}
\caption{Attenuation across accumulation kernels. Share of the steady-state slope recovered against hold length, for five accumulation kernels. Left: the last observation of the block, which recovers $F_L$. Right: the block average, the estimator used in the cost calculation, which recovers $G_L$. Points are simulated, lines are theory. Only the finite-memory kernel reaches one at a finite hold.}
\label{fig:dynamic}
\end{figure}

A further Monte Carlo exercise addresses the result that changes how such a study should be run. A stock with persistence $a$ is driven by blocks of constant deployment of length $L$, and the arm contrast is computed either from the last observation of each block or from the block average. \Cref{fig:dynamic} reports the share of the steady-state slope recovered.

\subsection{Validation: replication}

The second exercise checks the variance formulas against a panel generated with the estimated cross-sleeve correlation.

\begin{table}[t]
\centering
\caption{Variance of the arm contrast under the two assignment schemes}
\label{tab:replication}
\small
\begin{tabular}{@{}lrrrrr@{}}
\toprule
Sleeves $P$ & 10 & 20 & 50 & 100 & 400 \\
\midrule
Contemporaneous, simulated & 0.334 & 0.160 & 0.066 & 0.033 & 0.008 \\
\quad predicted $4(1-\bar\rho)/P$ & 0.326 & 0.163 & 0.065 & 0.033 & 0.008 \\
\midrule
Staggered, simulated & 0.534 & 0.436 & 0.395 & 0.399 & 0.394 \\
\quad predicted $2[1+(P-1)\bar\rho]/P$ & 0.535 & 0.453 & 0.405 & 0.388 & 0.376 \\
\midrule
Ratio & 1.6 & 2.7 & 6.0 & 12.1 & 48.2 \\
\bottomrule
\end{tabular}

\vspace{0.35em}
\begin{minipage}{0.94\textwidth}\footnotesize
6{,}000 periods, unit residual variance, $\bar\rho=0.186$. Under contemporaneous assignment the treated and control sleeves are drawn afresh within each period; under staggered assignment the arms occupy different periods. The contemporaneous variance falls as $1/P$ without limit; the staggered variance converges to $2\bar\rho=0.372$, so recruiting sleeves stops helping.
\end{minipage}
\end{table}

The prediction is exact in simulation. \Cref{tab:replication} generates a panel with the estimated cross-sleeve correlation and forms the contrast under each scheme. Contemporaneous assignment reproduces $4(1-\bar\rho)/P$ to three decimal places at every scale and continues to fall as sleeves are added. Staggered assignment reproduces $2[1+(P-1)\bar\rho]/P$ and flattens at $2\bar\rho$: from fifty sleeves onwards, further recruitment yields no measurable reduction. The ratio between the two schemes rises from $1.6$ at ten sleeves to $48$ at four hundred, which is the quantitative content of the design rule.

\subsection{The design results at the calibrated magnitudes}

With the inputs in hand the design results become a schedule: how long a study must run, at what scale, and what it gives up while it runs.

\begin{table}[t]
\centering
\caption{Calendar months required for 80\% power}
\label{tab:feasibility}
\small
\begin{tabular}{@{}lrrrrr@{}}
\toprule
Hold length $L$ & 1 & 6 & 12 & 24 & 48 \\
\midrule
Targets & impact & short & medium & near long run & long run \\
Not suited to & steady state & steady state & steady state & memory & memory \\
Recovery factor $G_L$ & 0.082 & 0.252 & 0.402 & 0.595 & 0.771 \\
\midrule
\multicolumn{6}{@{}l}{Block average, uncorrected, targeting the finite-hold effect}\\
$P=100$ & 26{,}565 & 2{,}841 & 1{,}112 & 509 & 302 \\
$P=400$ & 6{,}641 & 710 & 278 & 127 & 76 \\
\midrule
\multicolumn{6}{@{}l}{Block average, deattenuated with oracle weights, targeting the steady-state effect}\\
$P=100$ & 26{,}565 & 2{,}389 & 934 & 441 & 275 \\
$P=400$ & 6{,}641 & 597 & 233 & 110 & 69 \\
\midrule
Staggered inflation, $P=100$ & 23.9 & 23.9 & 23.9 & 23.9 & 23.9 \\
\bottomrule
\end{tabular}

\vspace{0.35em}
\begin{minipage}{0.94\textwidth}\footnotesize
$\sigma=3.61\%$ and $\bar\rho=0.186$ are read off the strategy panel, $a=0.9177$ and $g=c(1)-c(0)=0.123$ are transported under the bridges above at own share $\theta=1$; the implied $\kappa=c'(0)$ is $0.073$. At $\theta=\tfrac12$ and $\tfrac14$ every entry is multiplied by four and sixteen, so the two-year row at a hundred sleeves becomes $2{,}036$ and $8{,}144$ months. Computed under \Cref{as:linear} at $\Omega_D=4(\widehat S-\widehat C)/P$ with $\widehat S-\widehat C=11.01$. Both panels are evaluated at the transported kernel: the first targets the finite-hold effect and needs $G_L$ only to state the gap it can detect, the second targets the steady-state effect and needs $F$ and $\Sigma$ throughout. Neither is model free. The oracle rows use $v^{*}$ of \Cref{prop:optw} and are unattainable without knowing $F$; they bound from below what an implementable weighting could achieve, and the gain over the block average is sixteen per cent at a one-year hold and nine at a four-year one. \end{minipage}
\end{table}

The long-run variance is estimated directly rather than assumed. Since the contrast differences out whatever the sleeves share in a period, $\Omega_D=4(S-C)/P$ where $S$ is the long-run variance of one sleeve residual and $C$ the long-run covariance between two distinct sleeves. A Newey--West estimate at six lags on the panel gives $\widehat S-\widehat C=11.01$, against $\sigma^{2}(1-\bar\rho)=10.33$ for the same quantity under no serial dependence: allowing for autocorrelation raises the requirement by $6.5\%$, and by $13\%$ at a twelve-month bandwidth. The correction is therefore real but second order relative to the two design terms.

The assignment scheme is not second order. The mean pairwise correlation of the residuals is $\bar\rho=0.186$, so under staggered assignment the requirement is larger by the factor $[1+(P-1)\bar\rho]/(1-\bar\rho)$, which is $6.7$ at $P=25$, $23.9$ at $P=100$ and $92.4$ at $P=400$. Sleeves must be randomised within the same calendar periods, and a study that staggers them inherits the saturation.

\Cref{tab:feasibility} reports the schedule. Two conclusions follow, both about scale rather than patience. Under the central calibration and the fixed-horizon design considered here, a single-signal experiment at monthly frequency is not practically attainable: even ignoring attenuation, resolving a twelve-basis-point erosion against a residual volatility of $3.6\%$ requires 17{,}993 strategy-months. Replication resolves this only under contemporaneous assignment, where a hundred sleeves with two-year holds need forty-two years and four hundred with four-year holds six. Capacity is therefore a book-level estimand before it is a signal-level one.
Deattenuating under the wrong kernel biases the answer multiplicatively, and the size of that error can be measured. Generating the contrasts under alternative accumulation profiles and correcting as if the kernel were geometric at the calibrated persistence gives a bias of $+4\%$ under a mixture of two rates, $+30\%$ under a kernel of finite twelve-month memory, and $-50\%$ under a longer tail with $a=0.97$. The last is the case the persistence bridge is least able to exclude, and a factor of two is larger than any of the design choices the paper studies.

Calendar time is not what the experiment costs. Running an arm away from the scale the book would otherwise choose gives up edge, which is the quantity an institution would be asked to authorise. Write $V(\beta)=\beta\{\mu-c^{\mathrm{own}}(\beta)-c^{\mathrm{agg}}(\bar w)\}$ for deployed capital times the per-unit edge surviving at that deployment, holding aggregate deployment at its prevailing level, and $\beta^{*}=\arg\max V$ for the scale a sleeve would run absent the experiment. This is an own-sleeve calculation at a fixed aggregate, not a book-level one: the value of scaling the whole strategy carries the aggregate term as a function of $\beta$ rather than as a constant, and by \Cref{prop:tradeoff} that term is not identified here, so the book-level optimum is scenario-dependent and we do not compute it. At the calibration of \Cref{tab:calib}, $\beta^{*}=1.73$ and $V(\beta^{*})=0.739$ percentage points per month per unit of reference book, the additive aggregate term dropping out of both. A two-arm design with sleeves split evenly forgoes
\[
 \Lambda(\beta_0,\beta_1)=V(\beta^{*})-\tfrac12\{V(\beta_0)+V(\beta_1)\}
\]
per month for the $T$ months the power calculation requires, and the total is the product.

\begin{table}[t]
\centering
\caption{The experiment priced in forgone edge}
\label{tab:cost}
\small
\begin{tabular}{@{}lrrrrr@{}}
\toprule
Arms $\{\beta_0,\beta_1\}$ & Gap $g$ & Months & Forgone rate & Total forgone & Share of attainable \\
\midrule
$\{0,\,1\}$ & 0.123 & 509 & 0.449 & 228 & 0.61 \\
$\{1.5,\,2.0\}$ & 0.124 & 501 & 0.021 & 11 & 0.03 \\
$\{2.4,\,2.9\}$ & 0.169 & 270 & 0.348 & 94 & 0.47 \\
$\{1.5,\,3.0\}$ & 0.447 & 39 & 0.327 & 13 & 0.44 \\
$\{0,\,4\}$ & 1.092 & 6 & 1.517 & 10 & 2.05 \\
\bottomrule
\end{tabular}

\vspace{0.35em}
\begin{minipage}{0.94\textwidth}\footnotesize
A hundred sleeves, two-year holds, at the calibration of \Cref{tab:calib}. The forgone rate is $\Lambda$ in percentage points per month per unit of reference book, the total is $T\Lambda$ in cumulative percentage points undiscounted, and the share is $\Lambda/V(\beta^{*})$, which exceeds one when an arm destroys more edge than the optimal deployment creates.
\end{minipage}
\end{table}

\begin{figure}[t]
\centering
\includegraphics[width=.95\textwidth]{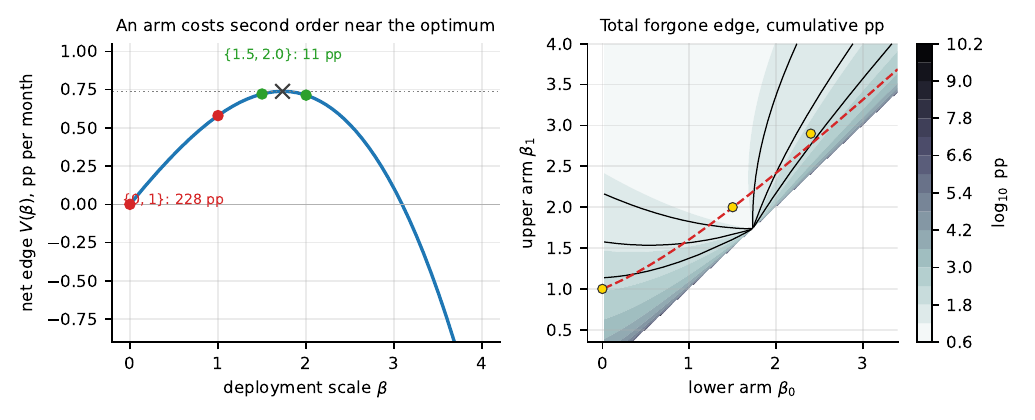}
\caption{The experiment priced in forgone edge. Left: net edge against deployment, with the arm pairs of \Cref{tab:cost} marked; because $V$ is flat at its maximum, an arm placed near $\beta^{*}$ costs second order in its distance while contributing first order to the identifying gap. Right: total forgone edge over the plane of arm pairs, with the equal-power contour at $g=0.123$ dashed, along which the cost varies by more than an order of magnitude.}
\label{fig:cost}
\end{figure}

\Cref{tab:cost} and \Cref{fig:cost} recast the headline number. The forty-two years quoted above are the cost of comparing a zero arm with a unit arm, which forgoes $228$ cumulative percentage points because the zero arm abandons the edge entirely. Arms at $1.5$ and $2.0$ need essentially the same duration, $501$ months, and forgo $11$: the same experiment at a twentieth of the cost, because both arms straddle the flat maximum. Widening to $\{1.5,3.0\}$ yields a gap almost four times larger and finishes in thirty-nine months for $13$ points. What makes an experiment expensive is therefore where its arms sit, not how long it runs, and the two are minimised by different designs. \Cref{prop:straddle} sets the floor. Narrow arms placed around the operating scale do not make the experiment cheap: at this calibration any symmetric pair around $\beta^{*}$ costs exactly $10.6$ cumulative points, because the second-order saving in forgone edge is offset by the squared first-order loss in the gap. Falling below that floor requires wide arms, $9.8$ points at $\{0,4\}$, and what limits the width is not cost. The identified set is never shorter than the grid spacing, so wide arms deliver detection and not resolution; and the linear approximation of \Cref{as:linear} degrades over a wide range of $\beta$. The design that follows is two-stage: wide arms to establish that erosion is detectable and to identify the kernel shape, then a grid dense enough around the crossing to resolve it.

The requirement is a scenario, and its sensitivity is concentrated in one input. A design that deattenuates with the transported $a_0=0.9177$ but faces a true persistence $a$ recovers the fraction $G_L(a)/G_L(a_0)$ of the steady-state effect, and the calendar requirement inflates by the square of its reciprocal. At two-year holds that fraction is $0.78$ if the truth is $a=0.946$ and $0.51$ if it is $a=0.97$, so the requirement rises from $509$ months to $841$ and to $1{,}981$; a much faster dissipation, $a=0.80$, would leave it at $259$. Nothing estimated on the panel moves the answer comparably, the serial dependence accounting for six per cent, and kernel shape is treated in \Cref{sec:compare}. The transported persistence is therefore what a study should try hardest to pin down, and \Cref{prop:traj} is what would pin it down from inside the experiment. That route has a price of its own: a hundred sleeves held two years at a time need eight years of calendar to place the persistence in $[0.39,0.99]$, sixteen to reach $[0.66,0.99]$ and sixty-four to reach $[0.82,0.98]$, the last still leaving the recovered effect uncertain by a factor from $0.74$ to $2.54$. Estimating the bridge to that precision is harder than detecting the slope. But the bridge that carries most of the conclusion is also the one a design can retire at low cost if it is asked the right question: \Cref{prop:traj} identifies the shape $F_j/F_L$ without any parametric restriction, and \Cref{tab:transport} finds the multi-horizon design a poor route to a point estimate of $c(\beta)$, not a poor route to $a$. A short multi-horizon pilot powered on the shape rather than on the level would therefore collapse the swing above before the expensive stage is committed, and the sequencing is itself the recommendation: estimate the kernel first, then transport nothing.

\subsection{The impact-model benchmark at the calibrated magnitudes}

The bound of \Cref{prop:impact} is one-sided but not vacuous, and how far it sits from the truth is a calibration question. Standard practice fits a square-root law in participation to the firm's own execution record, multiplies by turnover and horizon, and calls capacity the scale at which that cost consumes the gross edge: data the firm owns, millions of child orders rather than a few hundred monthly observations, and an estimate within a day.

\begin{table}[t]
\centering
\caption{How far an impact model overstates capacity}
\label{tab:impact}
\small
\begin{tabular}{@{}lrrr@{}}
\toprule
Crowding share of total erosion & 0.15 & 0.30 & 0.50 \\
\midrule
Impact-model capacity $\beta^{\mathrm{imp}}$ & 2.950 & 3.311 & 4.030 \\
Overstatement against $\bown=2.675$ & $+10\%$ & $+24\%$ & $+51\%$ \\
\bottomrule
\end{tabular}

\vspace{0.35em}
\begin{minipage}{0.94\textwidth}\footnotesize
At the calibration of \Cref{tab:calib} with $c^{\mathrm{crowd}}=\lambda c$ for the stated share, so that $\beta^{\mathrm{imp}}$ solves $(1-\lambda)c(\beta)=\mu-\tau_{\mathrm{out}}$. At $\lambda=0.50$ the implied capacity exceeds the mandate ceiling $\bar\beta=4$, which is itself informative: at that share an impact model reports the strategy as uncapped over the whole feasible range.
\end{minipage}
\end{table}

\Cref{tab:impact} quantifies the gap. A crowding share of fifteen per cent of total erosion puts the impact-model capacity ten per cent above the own-sleeve capacity; a half puts it beyond the mandate ceiling altogether. Which of these obtains is not knowable from execution data, which is the point: the impact model answers a question it can answer well and is silent on the one that separates a capacity estimate from a cost estimate. Two qualifications run the other way. The identification assumption is not innocuous, since order size is chosen by a trader with information, so an unadjusted impact model estimates a conditional association that may exceed or fall short of the causal execution cost; and if the model is fitted on realised net returns rather than on execution prices it will absorb part of the crowding channel and the direction of the bound is no longer guaranteed. The experiment and the impact model are therefore complements: one prices the cost of trading, the other the cost of the positioning that accompanies it.

\subsection{Which design to run}
\label{sec:compare}

Three design modules address two related objects, the steady-state response and the location of its crossing with the hurdle. The trajectory result says the accumulation profile is identified without a kernel and the sharpness result says a finite grid delivers an interval; neither says what either fact is worth once observations are finite. This section attaches a precision statement to each and compares the three on the same generated data.

Successive differences of the trajectory identify the kernel increments up to a common scale, so cohorts assigned contemporaneously to several hold lengths estimate the accumulation profile rather than importing it; \Cref{app:support} states this and prices it. Splitting a fixed budget over more cohorts raises the standard error of each increment while shrinking what each increment measures, and a budget of $120{,}000$ sleeve-periods over a longest hold of two years bears the trade-off out. Two cohorts leave $3{,}333$ blocks each and recover the profile with a root mean squared error of $0.056$; three, four, six and eight cohorts leave $2{,}500$, $2{,}000$, $1{,}429$ and $1{,}111$ blocks and give errors of $0.086$, $0.105$, $0.136$ and $0.162$. The finer resolution costs more precision than it delivers, so a multi-horizon design should use the coarsest split that separates the horizons it needs.

\begin{table}[t]
\centering
\caption{Transporting the kernel against estimating it, at equal resource}
\label{tab:transport}
\small
\begin{tabular}{@{}lrrrrrr@{}}
\toprule
& \multicolumn{3}{c}{Fixed horizon, transported} & \multicolumn{3}{c}{Multi horizon, estimated} \\
\cmidrule(lr){2-4}\cmidrule(lr){5-7}
True scale of accumulation & bias & s.d. & RMSE & bias & s.d. & RMSE \\
\midrule
Geometric, as transported & $+0.000$ & 0.003 & 0.003 & $-0.009$ & 0.057 & 0.058 \\
Two speeds & $-0.007$ & 0.003 & 0.008 & $+0.030$ & 0.010 & 0.032 \\
Finite memory & $-0.037$ & 0.003 & 0.037 & $-0.009$ & 0.012 & 0.014 \\
\bottomrule
\end{tabular}

\vspace{0.35em}
\begin{minipage}{0.96\textwidth}\footnotesize
600 replications, target $-c(1)=-0.123$, all quantities in percentage points per month. Both designs receive the same budget of 120{,}000 sleeve-periods, so a cohort of hold $L$ costs $L$ periods per block; the transported kernel is geometric at $a=0.9177$ whatever the truth. The fixed-horizon design has the lower root mean squared error under the geometric and two-speed kernels, the multi-horizon design under finite memory.
\end{minipage}
\end{table}

\Cref{tab:transport} settles the comparison, and it does not settle it uniformly. Compared on the same budget of sleeve-periods and on root mean squared error, transporting the kernel wins when the transported form is close to the truth and loses when it is badly wrong. If the truth is the geometric kernel that is transported, the fixed-horizon design has an error an order of magnitude smaller; if the truth mixes two speeds, it still wins by a factor of four, because the bias a misspecified geometric induces is smaller than the noise of learning the kernel from a single trajectory. Under finite memory the ranking reverses: the transported geometric carries a bias of $0.037$ against a target of $0.123$, three tenths of it, which the multi-horizon design avoids at a cost in variance that is smaller still. The design question is therefore not which route is better in general but how far the bridge of the bridges above is likely to be from the truth, and a multi-horizon design is worth its cost precisely when that distance is large.

The design implication is the opposite of the one the identification result invites. A multi-horizon design is worth running for the kernel shape, which \Cref{prop:traj} identifies robustly and \Cref{prop:multiprec} prices, and not as a route to a sharper point estimate. The bridge is a liability in principle and an asset in practice at this sample size, and the response is to report the terminal bound, which needs no kernel, alongside whichever point estimate is given.

\begin{table}[t]
\centering
\caption{Capacity set: placement rule against reporting rule}
\label{tab:designs}
\small
\begin{tabular}{@{}llrrr@{}}
\toprule
Placement & Reporting & Coverage & Length & In units of $r$ \\
\midrule
Fixed grid & bracket between point estimates & 0.479 & 0.284 & 0.85 \\
Fixed grid & simultaneous band set & 1.000 & 2.386 & 7.12 \\
Adaptive & bracket between point estimates & 0.514 & 0.371 & 1.11 \\
Adaptive & simultaneous band set & 1.000 & 2.774 & 8.27 \\
\bottomrule
\end{tabular}

\vspace{0.35em}
\begin{minipage}{0.96\textwidth}\footnotesize
4{,}000 replications. Arms are drawn from a candidate set of $41$ equally spaced points on $[0,\bar\beta]$ fixed before the experiment, eleven of them for the fixed grid and up to eleven for the adaptive rule, which stops when the surviving interval admits no new candidate. Nominal coverage is $0.90$ by a Bonferroni band simultaneous over all $41$ candidates, which is what \Cref{prop:band} requires when the locations are data-selected. True own-sleeve capacity $2.675$ with slope $-0.340$ there, so $r=0.335$.
\end{minipage}
\end{table}

\Cref{tab:designs} separates the two decisions. The reporting rule determines validity: a bracket between point estimates covers the truth less than half the time on a fixed grid and less than a fifth under sequential refinement, because each stage conditions on a bracket that may already be wrong, while the band set is valid by \Cref{prop:band} in both cases. The placement rule then determines length, and the ordering is the reverse of what might be expected. Adaptive refinement returns a longer set than a fixed grid of the same size, $8.3$ multiples of the resolution scale against $7.1$, because an arm whose interval straddles the hurdle can never be an endpoint of the reported set: concentrating arms near the crossing spends them exactly where the band cannot use them. The fixed grid is the better placement rule here, and the recommendation is to place arms on a grid, report the band set, and never report the bracket. Both band sets remain several multiples of $r$, the price of correcting over the whole candidate set, and a multiplier bootstrap would shorten them without losing validity.

Two groups of natural experiments have been proposed as substitutes for randomising deployment. The first, publication events, is already used in the calibration above. The second requires a defensible exclusion restriction. Securities-lending supply shifts \citep{duffie2002,cohen2007,saffi2011} and index reconstitutions move deployment for reasons plausibly unrelated to a signal's own edge. Neither is automatically valid: lending supply enters the net return directly through the borrow cost, and a reconstitution changes demand, liquidity and prices at once. Each requires an argument that the instrument reaches the adjusted return only through deployment, and a reported first stage.

\Cref{app:protocol} sets out the commitments to be fixed in advance.

\section{Conclusion}

The design follows from primitives rather than from convenience, and three consequences follow for practice.

Three consequences follow for practice. The estimand depends on how long the assigned level is held, so hold length is a modelling choice rather than a tuning parameter, and the shortest hold targets the impact-period effect rather than the steady-state slope. The hold that best identifies that slope is the one that best conceals whether erosion carries memory, so a single design cannot answer both questions. And replication across sleeves recovers its full value only when treated and control sleeves are contrasted within the same calendar period.

Two of the results point away from the methods they suggest. At our calibration optimal weighting improves on the block average by seven per cent and an estimator that learns the kernel from the trajectory it deattenuates loses far more than that, so the block average deattenuated with a transported kernel survives as the working estimator; and a sharp identified set is not a confidence set, so where arms are placed and what is reported are separate decisions.

Priced at the calibrated magnitudes and under a linear erosion map, the resulting experiment is expensive but finite. It is out of reach for a single strategy and within reach only for an institution able to randomise many sleeves simultaneously. That explains why the capacity of a trading strategy has been inferred observationally rather than measured through controlled variation in deployment, and it identifies the one design feature, contemporaneous assignment across sleeves, that separates an experiment of finite length from one whose requirement does not fall with scale.

Two limits should travel with the numbers. The calendar figures rest on transported bridges that the panel cannot identify and that the experiment could estimate only at a cost exceeding the one it pays to detect the slope, so declaring the bridge costs less than removing it; and because parallel sleeves trade the same securities, what a within-date contrast identifies is the effect of one sleeve's own assignment at a fixed average deployment. That is not a defect of this design but a property of within-date variation: an unrestricted calendar component absorbs the aggregate crowding exactly, so no estimator recovers it. Capacity experimentation fails when the estimand is left implicit; once it is declared, the design rules are explicit.

\bibliographystyle{apalike}
\bibliography{references,paperB_refs,newrefs}

\appendix

\section{The observational benchmark}
\label{app:boundary}

Write the adjusted return as $A_t=\mu-h+\varepsilon_t$ and suppose its law depends on $(\mu,h)$ only through the net edge $e=\mu-h$, each one-period mean equalling $e$. Two pairs then generate the same law for every $T$ if and only if they share the net edge: sufficiency is immediate, and for necessity equal laws have equal one-period means, which the normalisation identifies with the net edges. The fiber through a law with net edge $e$ is $\{(\mu,h):\mu-h=e\}$, so neither component is point identified, and equality of measures gives equality of expectations, so power equals size against the paired alternative. Bounding feasible erosion by $\bar h$ puts $\mu$ in $[e,e+\bar h]$ with every point attained, so the set is sharp with width $\bar h$ and more observations shrink uncertainty about $e$ without shrinking it. This is the benchmark the design is priced against.

\section{Supporting results}
\label{app:support}

The results below support the design results in the text without being design results themselves: the first two fix the class of erosion dynamics, the third collects the standard linear-model conditions whose conjunction constrains the design, and the last two price the multi-horizon variant.

\begin{proposition}[Normalisation within the scalar linear Markov class]\label{prop:charac}
\Cref{as:linear} admits exactly one functional form, indexed by the persistence $a$ and the scale response $c$:
\begin{equation}
 h_{t+1}=a\,h_t+(1-a)\,c(W_t),\qquad a\in[0,1).
 \label{eq:stock}
\end{equation} 
\end{proposition}

\begin{proposition}[Kernel representation]\label{prop:kernel}
Replace (iii) by memory of arbitrary finite or infinite order and retain (i), (ii), (iv) and (v). The structure is then of Hammerstein type, a static nonlinearity followed by a linear dynamic filter, $x_t=c(W_t)$ and $h_t=\mathcal L(\{x_{t-s}\})$, with $\mathcal L$ causal, linear, stable and time invariant in the transformed input $x$, not in deployment; add monotonicity, so that an additional unit of deployment does not reduce erosion at any horizon. Then the erosion state admits the representation
\begin{equation}
 h_t=\sum_{s\geq0}\omega_s\,c(W_{t-s}),
 \qquad \omega_s\geq0,\qquad \sum_{s\geq0}\omega_s=1,
 \label{eq:kernel}
\end{equation}
and the pair $(c,\{\omega_s\})$ is unique. \Cref{eq:stock} is the special case $\omega_s=(1-a)a^{s}$. 
\end{proposition}

\begin{proposition}[Identification, cancellation, efficiency]\label{prop:rank}
Let $H(Z_t)$ be a design matrix built from the assigned scales, either the indicators of the $J$ non-reference arms or a basis for the scale response, so that $Y_t=H(Z_t)\tau+X_t\theta+u_t$ with $\tau$ the vector of arm effects; a single coefficient would not carry a nonlinear scale response across several arms. Let $X_t$ hold the sleeve and date effects and any further nuisance regressors and $M_X$ be the residual maker of $X$.
\begin{enumerate}[label=(\roman*),leftmargin=*]
\item Identification. $\tau$ is identified if and only if $H(Z)^{\top}M_XH(Z)$ has full rank, and a linear functional $a^{\top}\tau$, such as the contrast between two specified arms, is identified if and only if $a$ lies in the row space of $M_XH(Z)$, equivalently $a\perp\ker M_XH(Z)$. A non-zero column is not sufficient: two columns can both survive residualisation and still be collinear, leaving combinations of their coefficients unidentified. If every sleeve receives the same scale on a given date, every column of $H(Z)$ lies in the span of the date effects and no arm effect is identified.
\item Cancellation. With a common date component, $Y_t=\delta_t\iota+\tau Z_t+u_t$, a contrast $q^{\top}Y_t$ is free of $\delta_t$ if and only if $q^{\top}\iota=0$. The condition places no restriction on $\delta_t$, so it covers arbitrary and possibly non-stochastic date effects; the weights satisfying it are not unique.
\item Efficiency. Among contrasts with $q^{\top}\iota=0$ that load unit weight on the arm effect of interest, the minimum-variance choice is $q^{*}=\Sigma^{-1}C(C^{\top}\Sigma^{-1}C)^{-1}c$, with $C=[H(Z),X]$ and $c$ the vector selecting the arm effect of interest while annihilating the rest. Under equal variances, equicorrelation and balanced arms it is the simple difference of arm means.
\end{enumerate} 
\end{proposition}

\begin{corollary}[Multi-horizon design]\label{cor:multi}
Successive differences of the trajectory identify the kernel increment at each observed horizon,
\[
 \E[D_{j+1}]-\E[D_j]=-c(\beta)\,\omega_j ,\qquad j<L,
\]
so a design that records the path, or that assigns cohorts to hold lengths $L_1<\dots<L_K$ within the same calendar periods, identifies $(\omega_0,\dots,\omega_{L_{\max}-1})$ up to the common scale $c(\beta)$. The unobserved tail $\sum_{s\geq L_{\max}}\omega_s$ is not identified and must be bounded, assumed, or made negligible by a long enough horizon. A multi-horizon design therefore delivers a steady-state point estimate only once that tail is closed, and the exercises below close it by fitting a geometric form to the observed shape; the row labelled multi horizon in \Cref{tab:transport} should be read as multi horizon plus that tail assumption. Cohorts must overlap in calendar time, since staggering them across dates reintroduces the common component that \Cref{prop:rank} requires the design to difference out. 
\end{corollary}

\begin{proposition}[Resolution against precision in a multi-horizon design]\label{prop:multiprec}
Assign cohorts contemporaneously to holds $L_1<\dots<L_K$ at a common scale, the design of \Cref{cor:multi},, with a total budget of $B$ block-observations split evenly, and let $\Omega$ be the long-run variance of a single block contrast. The budget is measured in sleeve-periods, so a cohort of hold $L_k$ with $n_k$ blocks costs $n_kL_k$ and $B=\sum_kn_kL_k$. Allocate a common number of blocks to every cohort, $n_k=n$ for all $k$, which is the rule the exercise below implements; the budget then fixes
\[
 n=\frac{B}{\sum_{k\le K}L_k},
 \qquad\text{so}\qquad
 \Var(\widehat D_k)=\frac{\Omega}{n}=\frac{\Omega\sum_{k\le K}L_k}{B}
\]
for every cohort. Cohort $k$ delivers $\widehat D_k$ with mean $-c(\beta) F_{L_k}$ and that variance. The normalised shape $\widehat D_k/\widehat D_K$ estimates $F_{L_k}/F_{L_K}$, and by the delta method the estimated increment between adjacent horizons has standard error of order
\[
 \sqrt{\frac{\Omega\sum_{k\le K}L_k}{B}}\cdot\frac{1}{|c(\beta)|F_{L_K}}\bigl(1+o(1)\bigr),
\]
which increases in $K$ at a fixed budget because refining a nested grid of horizons up to $L_K$ raises $\sum_kL_k$ roughly in proportion to $K$, while the increments being resolved sum to $F_{L_K}$ whatever $K$ and so become individually smaller. Under an even split over a nested refinement of a fixed range, precision per increment therefore falls in $K$ on both counts. Whether the extra temporal resolution repays that loss depends on the kernel and on the objective, and the budget exercise below shows that at this calibration it does not.
\end{proposition}

\section{Derivations}
\label{app:proofs}

The identification of the finite-hold curve at the assigned arms is the g-computation identity of \citet{robins1986}, used without restatement. Of the results in the text, two are not applications of an existing framework and are proved; the rest follow from the framework named where they are stated, or are algebra, and are derived here in the order the results appear.

\subsection{Derivation of \texorpdfstring{\Cref{prop:sharp}}{the sharp identified set for capacity}}

Containment follows from the intermediate value theorem, and the exclusions from continuity as stated. For sharpness, fix $b$ in the stated set and interpolate between the arms by any continuous non-increasing curve with $m(b)=\tau_{\mathrm{out}}$: descend from $m(\beta_j)$ to $\tau_{\mathrm{out}}$ on $[\beta_j,b]$ and from $\tau_{\mathrm{out}}$ to $m(\beta_{j+1})$ on $[b,\beta_{j+1}]$. Such a curve exists because $m(\beta_j)\geq\tau_{\mathrm{out}}>m(\beta_{j+1})$, it agrees with the identified values at every arm, and its capacity is exactly $b$. No point of the stated set can therefore be excluded.

\subsection{Derivation of \texorpdfstring{\Cref{prop:band}}{a capacity set with coverage}}

On the event that the band covers the mean of every point of $\mathcal G$, which has probability at least $1-\alpha$ and does not depend on which points the rule selects, $\widehat m(\beta_j)-c\,\widehat{\mathrm{se}}_j\geq\tau_{\mathrm{out}}$ implies $m(\beta_j)\geq\tau_{\mathrm{out}}$ and hence $\bown\geq\beta_j$ by monotonicity, so the lower endpoint is a valid lower bound; symmetrically for the upper. That event has probability at least $1-\alpha$ by simultaneity, and the argument conditions on the arms actually assigned, so sequential placement does not affect it.

\subsection{Proof of \texorpdfstring{\Cref{prop:own}}{what a sleeve-level contrast identifies}}

By exposure consistency the outcome depends on $Z_b$ only through $e_p(Z_b)$. \Cref{as:overlap} fixes $\sum_qZ_{qb}$, hence $\bar W_b$, across every vector in the support of the design, so two sleeves in the same block have exposures differing only in the first coordinate, by $\Delta$. Applying \Cref{prop:kernel} to each component of the erosion recursion and differencing gives $\mathrm{CDE}(\beta,\beta_0)R_L(u)$, with no residual term. For the last claim, $\mathrm{TE}^{\mathrm{strat}}$ requires exposures whose second coordinates differ by $\Delta$, and the design distribution places no mass on such pairs.

\subsection{Proof of \texorpdfstring{\Cref{prop:tradeoff}}{when aggregate crowding is identified within a date} and of \texorpdfstring{\Cref{cor:zerosum}}{zero-sum contrasts under homogeneous overlap}}

For (i), a common row $\gamma^{\top}$ gives $\gamma_p^{\top}W_t=\gamma^{\top}W_t$ for every $p$, so $g_t=c^{\mathrm{agg}}(\gamma^{\top}W_t)\iota$ enters each $Y_{pt}$ additively and identically. Replacing $\mu_t$ by $\mu_t-c^{\mathrm{agg}}(\gamma^{\top}W_t)$ and $c^{\mathrm{agg}}$ by the zero function leaves every $Y_{pt}$ unchanged, and $\mu_t$ is unrestricted so the replacement is admissible. For the corollary, $q^{\top}\{\mu_t\iota\}=\mu_t(q^{\top}\iota)$ vanishes for every $\mu_t$ if and only if $q^{\top}\iota=0$, and that condition annihilates $c^{\mathrm{agg}}(\bar W_t)\iota$ too. For (ii), $\iota^{\perp}$ and $g_t^{\perp}$ are hyperplanes, and one hyperplane contains another only if they coincide, which would require $g_t\in\operatorname{span}(\iota)$; any $q\in\iota^{\perp}\setminus g_t^{\perp}$ therefore serves.

\subsection{Derivation of \texorpdfstring{\Cref{prop:multiprec}}{resolution against precision in a multi-horizon design}}

Under a common number of blocks per cohort, $n=B/\sum_kL_k$, so each cohort mean has variance $\Omega\sum_kL_k/B$, which grows with $K$ along a nested refinement. Writing $R_k=\widehat D_k/\widehat D_K$, the delta method gives $\Var(R_k)\approx(\Omega\sum_kL_k/B)(1+R_k^2)/(c(\beta) F_{L_K})^{2}$, and the increment is a difference of two such ratios. Writing $R_k=\widehat D_k/\widehat D_K$, the delta method gives $\Var(R_k)\approx(\Omega\sum_kL_k/B)(1+R_k^2)/(c(\beta)F_{L_K})^{2}$, and the increment is a difference of two such ratios. The increments sum to a fixed total, so each is smaller the finer the grid. If the cohorts share calendar dates the ratios are correlated and the variance carries the additional term $-2R_k\Cov(\widehat D_k,\widehat D_K)$, which the contemporaneous design makes non-zero; the exercise below estimates the resulting error directly rather than through the approximation.

\subsection{Derivation of \texorpdfstring{\Cref{prop:straddle}}{the local opportunity cost of an experiment}}

Expanding about $\beta^{*}$ with $V'(\beta^{*})=0$ gives $V(\beta^{*}\pm d)=V(\beta^{*})+\tfrac12V''(\beta^{*})d^{2}+O(d^{3})$, and the odd terms cancel in the average, so $\Lambda=V(\beta^{*})-\tfrac12\{V(\beta^{*}-d)+V(\beta^{*}+d)\}=-\tfrac12V''(\beta^{*})d^{2}+O(d^{4})$. The gap is $2c'(\beta^{*})d+O(d^{3})$. Hence $C=K\Lambda/\{c(\beta_1)-c(\beta_0)\}^{2}\to K\{-V''(\beta^{*})\}/\{8c'(\beta^{*})^{2}\}$. Under $c(\beta)=\kappa\beta+\zeta\beta^{2}$ and $V(\beta)=\mu\beta-\kappa\beta^{2}-\zeta\beta^{3}$ the expansions terminate: $\Lambda=-\tfrac12V''(\beta^{*})d^{2}$ and the gap is $2c'(\beta^{*})d$ exactly, so $C=C^{*}$ for every $d$.

\subsection{Derivation of \texorpdfstring{\Cref{prop:impact}}{what an impact model identifies}}

Since $c^{\mathrm{crowd}}\geq0$ we have $c\geq c^{\mathrm{exec}}$ pointwise, so $\mu-c^{\mathrm{exec}}\geq\mu-c$ everywhere. Both are decreasing, and $\bown$ is the crossing of the smaller with $\tau_{\mathrm{out}}$, so the larger crosses weakly later and $\beta^{\mathrm{imp}}\geq\bown$. Equality of the two crossings forces $c^{\mathrm{crowd}}(\bown)=0$, and since $c^{\mathrm{crowd}}$ is non-negative, increasing and zero at zero, that extends to the whole of $[0,\bown]$.

\subsection{Derivation of \texorpdfstring{\Cref{prop:outcome}}{total effect, direct effect, and the adjustment convention}}

Write $\varphi=\E[F]$ and note that randomisation makes the law of $F$ identical across arms.
(i) $\E[R(\beta)]-\E[R(0)]=-c(\beta)+(b_0+\lambda\beta)^{\top}\varphi-b_0^{\top}\varphi=-c(\beta)+\lambda^{\top}\varphi\beta$.
(ii) For an arm-invariant $c$, the adjusted contrast subtracts $c^{\top}\E[F]$ from both arms, which cancels; hence the estimand is unchanged. The variance of the arm mean of $R-c^{\top}F$ is $(b-c)^{\top}\Sigma(b-c)+\sigma^2$ with $\Sigma=\Var(F)$, minimised over arms at $c=\E[b(W)]$.
(iii) With the loading estimated inside arm $\beta$, the adjusted arm mean converges to $\E[R(\beta)]-b(\beta)^{\top}\varphi=\mu-c(\beta)$; differencing across arms gives $-c(\beta)$.
(iv) Subtracting (iii) from (i) gives $\lambda^{\top}\varphi\beta$, which vanishes for all $\beta$ if and only if $\lambda^{\top}\varphi=0$. \qed

\subsection{Derivation of \texorpdfstring{\Cref{prop:feasibility}}{required scale under stock attenuation and correlated sleeves}}

Let $\varepsilon_{it}$ be the period residual of sleeve $i$, with variance $\sigma^2$ and average pairwise correlation $\bar\rho$. Write the within-period contrast as $\widehat D_t=a^{\top}\varepsilon_t$ with $a_i=+2/P$ on the treated half and $-2/P$ on the control half, so that $\sum_i a_i=0$. Then $\Var(\widehat D_t)=a^{\top}\Sigma a$ with $\Sigma=\sigma^2[(1-\bar\rho)I+\bar\rho\,\iota\iota^{\top}]$ under equicorrelation. Because $\iota^{\top}a=0$, the rank-one term drops out and
\[
 \Var(\widehat D_t)=\sigma^2(1-\bar\rho)\,a^{\top}a=\sigma^2(1-\bar\rho)\,\frac{4}{P},
\]
which is the stated expression; the same result holds for a general $\Sigma$ with $a$ drawn by balanced randomisation, on replacing $\bar\rho$ by the average off-diagonal correlation. Averaging the block-average outcome over $T/L$ blocks, each of which averages $L$ serially independent periods, gives $\Var=4\sigma^2(1-\bar\rho)/(PT)$, and setting the achieved gap $G_Lg$ equal to $c_{\alpha,\pi}$ standard errors yields $T$. Under staggered assignment $\iota^{\top}a\neq0$, the rank-one term survives, and the variance of the arm mean is $\sigma^2[1+(P-1)\bar\rho]/P$, which converges to $\sigma^2\bar\rho>0$. \qed

\subsection{Derivation of \texorpdfstring{\Cref{prop:hysteresis}}{what the ramp identifies}}

(i) If $h_t=g(W_t)$, the mean adjusted return during a hold at level $\beta_j$ is $\mu-g(\beta_j)$ regardless of the leg, so each term of $L$ has mean zero.

(ii) Under \cref{eq:stock} write $u_t=h_t-c(\beta_j)$ for the deviation during a hold at level $j$; then $u_{t+1}=a u_t$ and the mean deviation over $\ell$ periods starting from entry deviation $u_0$ is $u_0\,a(1-a^{\ell})/(\ell(1-a))$. Entering level $j$ from below along the increasing leg gives $u_0=-\{c(\beta_j)-c(\beta_{j-1})\}+o(1)$, and entering from above along the decreasing leg gives $u_0=+\{c(\beta_{j+1})-c(\beta_j)\}+o(1)$, where the remainder collects the residual deviation carried from earlier levels; on a regular grid both are $c'(\beta_j)\Delta\beta+o(\Delta\beta)$ by differentiability, and they coincide only to that order when $c$ is nonlinear. Since the adjusted mean is $\mu-c(\beta_j)-\bar u$, the difference between legs is $2c'(\beta_j)\Delta\beta\,a(1-a^{\ell})/(\ell(1-a))+o(1)$, which is $O(1/\ell)$ for $\ell$ large.

(iii) With $P$ sleeves and $\ell$ periods per level, the standard error of each matched difference is proportional to $1/\sqrt{P\ell}$, so the standardised statistic is proportional to $\sqrt{P\ell}\cdot(1-a^{\ell})/\ell\propto\sqrt P(1-a^{\ell})/\sqrt{\ell}$. This vanishes as $\ell\to\infty$ and as $\ell\to0$ relative to the sampling requirement, hence is maximised at an interior $\ell$. \qed

\subsection{Derivation of \texorpdfstring{\Cref{prop:charac}}{normalisation within the scalar linear Markov class}}

By (i)--(iv) the state satisfies $h_{t+1}=a h_t+b\,c(W_t)$ for constants $a,b$ with $|a|<1$; stability and nonnegativity give $a\in[0,1)$, the lower endpoint being the memoryless case. Under constant $W_t=\beta$ the fixed point solves $h^{*}=ah^{*}+b\,c(\beta)$, so $h^{*}=b\,c(\beta)/(1-a)$. Requirement (v) forces $b=1-a$. The pair $(a,c)$ is not otherwise restricted, so the primitives determine the form and leave those two objects free. The primitives pin the recursion down to the persistence $a$ and the scale response $c$ and no further, and relaxing the Markov order leaves a unique normalised accumulation kernel; \Cref{app:support} states both and proves them. What matters here is what they leave free. What they remove is freedom of functional form, and the two objects that remain are precisely the two the design has to learn or transport. Two objects are therefore separated that a single coefficient would conflate: $c$ governs how erosion varies with the size of the position, and $a$ how it builds up over time. The linear case $c(\beta)=\kappa\beta$ is the one in which marginal crowding is constant; the working model used for the simulations, $c(\beta)=\kappa\beta+\zeta\beta^{2}$, has increasing marginal crowding and belongs to the same family.

\subsection{Derivation of \texorpdfstring{\Cref{prop:kernel}}{kernel representation}}

Because $\mathcal L$ is causal, linear, stable and time invariant in $x$, it has a unique absolutely summable impulse response $\{\psi_s\}$ with $h_t=\sum_s\psi_sx_{t-s}$, and monotonicity gives $\psi_s\geq0$. Under constant deployment $\beta$ the input is constant at $c(\beta)$ and the steady state is $\bigl(\sum_s\psi_s\bigr)c(\beta)$, so requirement (v) forces $\sum_s\psi_s=1$ and $\omega_s=\psi_s$.

\subsection{Derivation of \texorpdfstring{\Cref{prop:optw}}{oracle minimum-variance recovery of the steady-state effect}}

For uniqueness, suppose $(c,\{\omega_s\})$ and $(\tilde c,\{\tilde\omega_s\})$ both satisfy the representation with weights summing to one and generate the same $h$ for every input path. Constant inputs give $c(\beta)=\tilde c(\beta)$ for every $\beta$, so the static parts agree; the dynamic parts then agree by uniqueness of the impulse response of a linear map. It is the normalisation that delivers this: without it, scaling $c$ by any positive constant and the weights by its reciprocal leaves $h$ unchanged. Substituting $\omega_s=(1-a)a^{s}$ recovers \cref{eq:stock}.

\subsection{Derivation of \texorpdfstring{\Cref{prop:rank}}{identification, cancellation, efficiency}}

Minimise $v^{\top}\Sigma v$ subject to $v^{\top}F=1$. The Lagrangian gives $2\Sigma v=\lambda F$, so $v\propto\Sigma^{-1}F$, and the constraint fixes the scale. The objective is strictly convex on the affine constraint set because $\Sigma\succ0$, so the minimiser is unique.

\subsection{Derivation of \texorpdfstring{\Cref{cor:multi}}{multi-horizon design}}

(i) is the Frisch--Waugh--Lovell characterisation applied to the block $H(Z)$: the least-squares coefficient vector exists and is unique precisely when $M_XH(Z)$ has full column rank, which fails when any column of $H(Z)$ lies in the span of $X$. For (ii), $q^{\top}(\delta_t\iota)=\delta_t(q^{\top}\iota)$ vanishes for every realisation of $\delta_t$ if and only if $q^{\top}\iota=0$. For (iii), minimising $q^{\top}\Sigma q$ subject to $C^{\top}q=c$ has the stated unique solution when $\Sigma\succ0$ and $C$ has full column rank, by the same Lagrangian argument as \Cref{prop:optw}.

\subsection{Derivation of \texorpdfstring{\Cref{prop:traj}}{trajectory identification and the one-sided bound}}

Immediate from $\E[D_j]=-c(\beta) F_j$ and $F_{j+1}-F_j=\omega_j$. Identification of the tail would require an observation at a horizon beyond $L_{\max}$, which the design does not provide.

(i) Ratios of the means eliminate $c(\beta)$ and identify $F_j/F_L$, whose differences give $\omega_j/F_L$. (ii) $F_L=\sum_{s<L}\omega_s\le\sum_{s\ge0}\omega_s=1$ by nonnegativity and the normalisation of \Cref{prop:kernel}, so $|c(\beta)|=|\E[D_L]|/F_L\ge|\E[D_L]|$. (iii) Any $F_L\in(0,1]$ consistent with the observed ratios yields an admissible kernel, so $c(\beta)$ is identified only up to the factor $1/F_L$.

\section{Path dependence: the ramp design and its power}
\label{app:hystprop}

The ramp visits levels $\beta_1<\dots<\beta_J$ upward and then downward, holding each for $\ell$ periods, and records the adjusted edge at every level on both legs. Write $d_j=\bar m^{\uparrow}_j-\bar m^{\downarrow}_j$ for the difference at level $j$ and $d=(d_1,\dots,d_J)^{\top}$.

\begin{proposition}[What the ramp identifies]\label{prop:hysteresis}
Under a memoryless erosion map, $\E[d]=0$ for every $\ell$ and every level grid, so the complete null of path invariance is $d=0$. Under \Cref{eq:stock} with $a>0$,
\[
 \E[d_j]\;=\;2\bigl\{c(\beta_j)-c(\beta_{j-1})\bigr\}\frac{a(1-a^{\ell})}{\ell(1-a)}\bigl(1+o(1)\bigr)
 \;=\;2c'(\beta_j)\,\Delta\beta\,\frac{a(1-a^{\ell})}{\ell(1-a)}\bigl(1+o(1)\bigr)
\]
along a regular grid with spacing $\Delta\beta$, so the expected difference is proportional to $1-G_\ell$, the transient component, while recovery of the steady-state capacity slope is proportional to $G_\ell$ itself. The two are complementary: a hold long enough to equilibrate the stock drives $1-G_\ell$ to zero and closes the loop the test relies on, and the hold that maximises power for path dependence is therefore shorter than the hold that minimises attenuation for the slope.
\end{proposition}

\Cref{tab:hysteresis} reports the comparison. Because the null is a vector, any scalar statistic is a projection $a^{\top}d$ and no projection is uniformly best. We compare three: the signed loop $\Lambda=\sum_j d_j$, the quadratic form $Q=d^{\top}\widehat S^{-1}d$, and the maximum $T_{\max}=\max_j|d_j|/\widehat{\operatorname{se}}(d_j)$. All three are referred to a sign-flip permutation distribution over sleeve-level loop vectors. That distribution is exact under the sharp null when the direction of the ramp is itself randomised across sleeves, so that reversing a sleeve's legs is a relabelling the design could have produced; with the direction fixed for every sleeve it is exact only under conditional symmetry of the sleeve-level differences, and is otherwise an approximation. We randomise the direction.

\begin{table}[t]
\centering
\caption{Rejection rates of three tests of path invariance}
\label{tab:hysteresis}
\small
\begin{tabular}{@{}lrrrr@{}}
\toprule
Alternative & Amplitude & Signed loop & Quadratic & Maximum \\
\midrule
Memoryless null & 0.00 & 0.040 & 0.045 & 0.050 \\
Dense, equal across levels & 0.10 & 0.210 & 0.117 & 0.113 \\
Sparse, one level & 0.10 & 0.235 & 0.443 & 0.520 \\
Cancelling, opposed halves & 0.30 & 0.113 & 0.723 & 0.555 \\
\bottomrule
\end{tabular}

\vspace{0.35em}
\begin{minipage}{0.94\textwidth}\footnotesize
400 replications, 60 sleeves, five levels on $[0,4]$, six-period holds, 300 sign-flips, $\sigma=3.611$. All three hold their size. The signed loop is best against the alternative its own projection matches and collapses to near its size against an alternative whose level-specific differences cancel, which is the case the vector null was introduced to cover. The quadratic form and the maximum are the appropriate choices against dispersed and concentrated departures respectively.
\end{minipage}
\end{table}

\section{The outcome definition as a choice of estimand}
\label{app:outcome}

Adjusting an outcome by a quantity the treatment itself moves is the situation \citet{rosenbaum1984} warns against, and it has a specific form here. Suppose deployment shifts exposures as well as the edge, $b_i(\beta)=b_{i,0}+\lambda_i\beta$, so that the adjusted return carries a factor-mediated channel $\lambda_i^{\top}\varphi\,\beta$ alongside the direct erosion.

\begin{proposition}[Total effect, direct effect, and the adjustment convention]\label{prop:outcome}
Under the exposure tilt with randomised arms sharing the same law of $F$:
\begin{enumerate}[label=(\roman*),leftmargin=*]
\item the total effect on the net return is $\mathrm{TE}(\beta)=-c(\beta)+\lambda^{\top}\varphi\,\beta$;
\item for any arm-invariant vector $c$, the contrast in $A=R-c^{\top}F$ also equals $\mathrm{TE}(\beta)$; the choice of $c$ affects only the variance, which is minimised at $c=\E[b(W)]$;
\item if the loading is estimated separately within each arm, the contrast equals $\mathrm{DE}(\beta)=-c(\beta)$, the direct effect that excludes the factor-mediated channel;
\item hence $\mathrm{TE}(\beta)-\mathrm{DE}(\beta)=\lambda^{\top}\varphi\,\beta$ equals the factor-mediated component, and the two coincide for all $\beta$ if and only if $\lambda^{\top}\varphi=0$.
\end{enumerate}
\end{proposition}

At the calibrated tilt the two estimands are $-0.088$ and $-0.123$, so the factor-mediated component is $0.035$, two fifths of the total effect in magnitude. A Monte Carlo exercise confirms it: the three arm-invariant specifications are indistinguishable from the total effect, the within-arm specification recovers the direct one, and fixing the loading before assignment reduces the standard deviation by about thirty per cent without moving the estimand, which is the variance reduction \citet{lin2013} analyses.

\section{Pre-registration protocol}
\label{app:protocol}

The following commitments should be fixed before an experiment begins. The estimand, stating whether the target is the capacity curve or the capacity itself and which hold length defines it. The arms, their allocation and the hold, with the hold justified against the anticipated dissipation half-life rather than by convention. The outcome, including the timestamp of every generated object and the adjustment convention, since the choice between an arm-invariant and a within-arm adjustment changes the estimand. How realised deployment will be measured, together with the first-stage strength below which the analysis will be reported as inconclusive. The inference, including the multiplier blocks matched to the outcome horizon. A commitment to report a one-sided bound when the curve does not cross the hurdle inside the feasible range, rather than extrapolating a fitted form. Whether the whole trajectory or only the block outcome will be recorded, since the first identifies the kernel shape and the second does not; a study that records it can estimate the dissipation rate rather than transporting it, which the panel used here cannot do. Which of the two answers will be reported, the model-free bound or the deattenuated point, with both required if the second is given. For a test of path dependence, the class of alternatives, since the vector null admits no uniformly best scalar. And the pooled scale and detectable gap implied by \Cref{prop:feasibility}, computed in advance, so that an underpowered study is recognised as such before it is run rather than after.

\section{Simulation details}
\label{app:sims}

All exercises use the panel of \Cref{sec:calib}: thirteen long--short strategies, eight common drivers screened from 125 candidates, loadings estimated by ordinary least squares over 2001:01--2023:07. Adjusted returns carry $\sigma=3.611$ percentage points per month, the stock persistence is $a=0.9177$, the scale response is $c(\beta)=0.073\beta+0.05\beta^{2}$ so that $c(1)=0.123$, and the exit hurdle is $\tau_{\mathrm{out}}=0.150$ against an uncrowded edge $\mu=0.703$. The exit hurdle and the curvature are declared design choices; every other constant is derived from the inputs.

Three accumulation kernels appear. The geometric kernel is $\omega_s=(1-a)a^{s}$ at $a=0.9177$. The two-speed kernel is the mixture $\omega_s\propto\tfrac12(1-\tfrac12)\tfrac12^{s}+\tfrac12(1-0.97)0.97^{s}$, a fast component with half-life one period and a slow one with half-life twenty-three, normalised to sum to one. The finite-memory kernel puts $\omega_s=1/12$ for $s<12$ and zero beyond. In every case $F_j=\sum_{s<j}\omega_s$ and $G_L=L^{-1}\sum_{j\le L}F_j$.

The feasible kernel estimator fits a geometric shape to the observed trajectory. Given block-averaged contrasts $\widehat D_1,\dots,\widehat D_L$, it forms the normalised profile $\widehat D_j/\widehat D_L$ and selects
\[
 \widehat a=\arg\min_{a\in\mathcal A}\sum_{j\le L}\Bigl\{\frac{\widehat D_j}{\widehat D_L}-\frac{1-a^{j}}{1-a^{L}}\Bigr\}^{2},
\]
over a grid $\mathcal A$ of 500 points equally spaced on $[0.05,0.995]$, and sets $\widehat F_j=1-\widehat a^{j}$. The same fit closes the unobserved tail in the multi-horizon exercise: the steady state is recovered as $\widehat D^{\mathrm{avg}}/\widehat G_L$ with $\widehat G_L$ computed from $\widehat a$, which extrapolates the geometric form beyond the longest horizon observed. That extrapolation is the tail assumption, and the row labelled multi horizon should be read as multi horizon plus it.

The recovery exercise uses two-year holds, a hundred sleeves split evenly between the two arms and forty blocks at design size, with the ten- and hundred-fold arms scaling both the sleeve count and the block count. The oracle uses the true $F$ and the sample covariance of the block contrasts; the feasible estimator uses $\widehat F$ from the fit above and the same sample covariance, ridged by $10^{-9}$ on the diagonal.

The multi-horizon exercise fixes a budget of $B=120{,}000$ sleeve-periods and a longest hold $L_K=24$. For each $K$ the horizon grid is $L_k=\mathrm{round}\{L_Kk/K\}$ for $k=1,\dots,K$, deduplicated, and every cohort receives the same number of blocks, $n=B/\sum_kL_k$; the resulting counts are reported with the exercise.

The capacity-set exercise draws arms from a candidate set fixed in advance. The fixed grid is $2+3S$ equally spaced points on $[0,4]$ for $S$ stages. The adaptive rule starts from $[0,4]$ and at each stage places three interior arms equally spaced within the surviving interval, together with its endpoints, then recomputes the interval from all arms assigned so far; it stops early if the interval becomes empty. The Bonferroni family is the maximum number of arms the rule can produce, $M=2+3S$, and the band half-width is $z_{1-\alpha/(2M)}\,\mathrm{se}$ with $\alpha=0.10$ and $\mathrm{se}=2\sigma/\sqrt{P\,n_{\mathrm{arm}}}$ at $P=100$ sleeves and $n_{\mathrm{arm}}=40$ block-observations per arm. Coverage is of $\bown(\bar w)=2.675$.

The bridge diagnostic proceeds as follows. The crowding state is the detrended log ninetieth percentile of implied borrow rates, and its innovation is the residual of an AR(1) fitted to that series. Each strategy's driver-adjusted return is regressed on that innovation at horizons $h=0,\dots,12$, giving $\widehat\theta_{i,h}$, and $(b_i,q)$ are chosen to minimise $\sum_{i,h}(\widehat\theta_{i,h}-b_iq^{h})^{2}$ with strategy-specific amplitudes and a decay common across strategies, over a grid of $q$ on $[0,0.995]$. The non-overlapping momentum family is the four strategies whose universes do not intersect: momentum in North America, Europe and Asia-Pacific, and industry momentum. Intervals come from 1{,}000 moving-block bootstrap replications with blocks of twelve months resampled to the original length. The exercise uses the 260 months on which all thirteen strategies and the borrow series overlap, and returns $\widehat q=0.280$ for the momentum family and $0.269$ for all thirteen, with intervals $[0.000,0.995]$ and $[0.000,0.965]$.

The publication check compares the mean adjusted return over the ten years before and after each strategy's publication date, taken from the original articles. To correct for the possibility that any ten-year window differs from its predecessor, the same statistic is computed at every admissible placebo date in that strategy's own history, and the raw change is reported net of the placebo median together with the percentile it occupies in the placebo distribution. The four strategies with identifiable dates give placebo-corrected declines of $-0.103$, $+0.576$, $-0.098$ and $+0.232$ percentage points per month for the book-to-market, industry momentum, profitability and investment strategies, at placebo percentiles $0.47$, $0.80$, $0.30$ and $0.72$.

For the outcome definition, $c(1)=0.123$, the exposure tilt is calibrated to $\lambda^{\top}\varphi=0.0348$ on the highest-premium driver, with unit perturbation, 600 observations per arm and 1{,}500 replications. The pre-treatment loading is estimated on the control arm. For the dynamic policies, blocks are generated from the stock recursion at each kernel with 600 blocks and 400 replications. For the ramp, five levels on $[0,4]$ are visited upward and then downward with six-period holds, 60 sleeves, 400 replications and 300 sign-flips, the direction of the ramp being randomised across sleeves. Replication counts and sample sizes are given with each table.

\end{document}